\documentclass[12pt,a4]{article}

\usepackage{a4}
\usepackage{amstext}
\usepackage{amsfonts}
\usepackage{amssymb}
\usepackage{color}
\usepackage{epsfig}
\usepackage{verbatim}

\newcommand{\gtapprox}{\raisebox{-0.5ex}{$\,\stackrel{>}{\scriptstyle\sim}\,$}}
\newcommand{\ltapprox}{\raisebox{-0.5ex}{$\,\stackrel{<}{\scriptstyle\sim}\,$}}

\newcommand{\be}{\begin{equation}}
\newcommand{\ee}{\end{equation}}
\newcommand{\ba}{\begin{eqnarray}}
\newcommand{\ea}{\end{eqnarray}}

\newcommand{\bfx}{{\bf x}}
\newcommand{\bfy}{{\bf y}}
\newcommand{\bfxo}{{\bf x}_0}
\newcommand{\bfzo}{{\bf z}_0}
\newcommand{\tr}{\textrm{Tr}}

\begin{document}


\begin{center}

{\large \bf On the definition and interpretation of a static quark 
anti-quark potential in the colour-adjoint channel}

\vspace{0.5cm}

\textbf{Owe Philipsen, Marc Wagner} \\
Goethe-Universit\"at Frankfurt am Main, Institut f\"ur Theoretische Physik, \\ 
Max-von-Laue-Stra{\ss}e 1, D-60438 Frankfurt am Main, Germany

\vspace{0.5cm}

May 25, 2013

\end{center}

\vspace{0.1cm}

\begin{tabular*}{16cm}{l@{\extracolsep{\fill}}r} \hline \end{tabular*}

\vspace{-0.4cm}
\begin{center} \textbf{Abstract} \end{center}
\vspace{-0.4cm}

We study possibilities to define a static quark anti-quark pair in a 
colour-adjoint orientation based on Wilson loops with generator insertions, 
using both lattice QCD and leading order perturbation theory in various gauges.
Non-perturbatively, the only way to obtain non-zero results while maintaining 
positivity of the Hamiltonian is by some form of temporal gauge.
In this case the correlator is equivalent to a gauge invariant
correlation function of a static quark anti-quark pair and a static 
adjoint quark, the resulting three-point potential is attractive. 
Saturating open colour indices with colour magnetic fields instead also leads to
a gauge invariant correlator. However this object is found to 
couple to the singlet sector only. There appears to be no lattice
observable that reproduces the repulsive adjoint potential predicted
by perturbation theory in Lorenz or Coulomb gauges.

\begin{tabular*}{16cm}{l@{\extracolsep{\fill}}r} \hline \end{tabular*}

\thispagestyle{empty}


\newpage

\setcounter{page}{1}

\section{Introduction}

Heavy quarkonium systems are most conveniently treated 
by means of effective theory methods like Non-Relativistic QCD 
(NRQCD) \cite{nrqcd} or potential Non-Relativistic QCD 
(pNRQCD) \cite{pnrqcd}. In such frameworks the 
heavy quark mass gets integrated out, which to leading order results in 
a static quark propagator, i.e.\ a temporal Wilson line.
The correlation function of a quark anti-quark pair at separation $r$
then factorises into a free propagator and the Wilson loop, whose 
exponential fall-off at large correlation time defines the static quark
anti-quark potential. 

Ever since the early treatments \cite{brown,svet,nad} there has been interest
in the colour-adjoint (or octet for $N=3$) channel, where the quark anti-quark
pair is in the adjoint of its product representation of $SU(N)$, 
$N\otimes \bar{N}=1 \oplus (N^2-1)$,
and hence the corresponding
mesonic states carry colour charge. While these are clearly ruled out as 
asymptotic states of the particle spectrum, they naturally appear as 
intermediate states in the framework of NRQCD, pNRQCD or in the presence of
a medium like the quark gluon plasma \cite{shur,pet}. A perturbative definition after
gauge fixing to Lorenz or Coulomb gauge appears to be straightforward and 
to leading order one finds
\cite{brown}
\begin{eqnarray}
\label{EQN790} (N^2-1) V^{T^a}(r) \ \ = \ \ -V^1(r) + \mathcal{O}(g^4)
\end{eqnarray}
($V^1$ denotes the singlet, $V^{T^a}$ the colour-adjoint static potential), where the 
next-to-leading order is also known \cite{kniehl}. However, the question of a 
non-perturbative generalisation is still open. 
Attempts to decompose Polyakov loop correlators
defined at finite temperature into singlet and adjoint channels 
\cite{svet,nad} were shown to fail non-perturbatively, 
since the exponential decay of 
both channels is dictated by the colour-singlet potential 
and the difference between them due to 
gauge-dependent matrix elements \cite{jp} (see also \cite{seiler}). 

In this paper we discuss the static quark anti-quark potential in the colour-adjoint channel at zero temperature based on the Wilson loop with generator insertions, 
as defined in section~\ref{SEC631}. This correlator is gauge-dependent and its non-perturbative evaluation requires gauge fixing. 
Lorenz gauges, which are commonly used in perturbation theory, are known to violate positivity and hence preclude the definition 
of a transfer matrix as well as a purely exponential decay of the correlator. 
Therefore, we mainly work with temporal gauge which preserves those properties. 
We study the resulting correlation functions by spectral analysis in terms of the transfer matrix (section~\ref{SEC632}) and compute some of them 
numerically for $SU(2)$ (section~\ref{SEC500}). We also discuss the use of Coulomb gauge, which 
allows for a well-defined transfer matrix too \cite{Zwanziger:1995cv}. However, non-perturbatively it requires a completion 
which can be implemented by a reduced form of temporal gauge.
On the other hand, saturating the open adjoint indices with
colour-magnetic fields, as suggested in the literature \cite{pnrqcd}, 
produces a 
gauge invariant observable but spectral analysis shows it to project on the colour-singlet
channel only.
In section~\ref{SEC118} we compare with leading-order perturbative calculations in the continuum using various gauges.

Our main results are:
\begin{itemize}
\item[(1)] In temporal gauge, the Wilson loop with generator insertions is equivalent to a gauge invariant quantity, where the 
generators are connected by an adjoint Schwinger line.

\item[(2)] Non-perturbatively, both in temporal as well as in ``completed Coulomb gauge'', the colour-adjoint static potential corresponds 
to a system of three static quarks, two fundamental and one adjoint quark, which form a colour-singlet state; 
this is in contrast to perturbatively calculated colour-adjoint static potentials in Lorenz or Coulomb gauge, where no adjoint quark is present.

\item[(3)] The non-perturbative colour-adjoint static potential is attractive (with even stronger binding than in the usual singlet channel), 
while the perturbative colour-adjoint static potential in Lorenz or Coulomb gauge is repulsive (cf.\ eq.\ (\ref{EQN790})).

\item[(4)] The repulsive colour-adjoint potential obtained in perturbation theory cannot be reproduced by any 
lattice observable considered here, not even at short distance.
\end{itemize}

Parts of this work have been presented at a conference \cite{Wagner:2012pk}.


\section{\label{SEC631}Static potentials based on Wilson loops}

The definition and calculation of the potentials between a static quark and 
anti-quark at distance $|\bfx -\bfy|$, each in the 
fundamental representation, is usually based on the trial states
\begin{eqnarray}
\label{EQN201} | \Phi^{\Sigma} \rangle \ \ \equiv \ \ 
\bar{Q}(\bfx) U^{\Sigma}(\bfx,\bfy) Q(\bfy) | 0 \rangle .
\end{eqnarray}
Here $Q$ and $\bar{Q}$ are static quark/anti-quark operators which are treated as spinless (one component) 
colour charges, since the spin decouples from the Hamiltonian in the static limit, 
\begin{eqnarray}
U^{\Sigma}(\bfx,\bfy) \ \ \equiv \ \ U(\bfx,\bfxo) \Sigma U(\bfxo,\bfy) ,
\end{eqnarray}
$\Sigma$ is an $N \times N$ matrix in the fundamental colour representation and 
$U(\bfx,\bfy)$ is a gluonic string represented by a 
straight spatial Wilson line connecting $\bfx$ and $\bfy$. 
Hence, $|\Phi^\Sigma\rangle$ is
a static meson with colour transformation properties 
determined by $\Sigma$.
For $\Sigma=1$ it is a colour-singlet, whereas for $\Sigma=T^a$, 
the generators of the
$SU(N)$ gauge group, it carries colour charge and 
transforms in the adjoint representation of $SU(N)$ at $\bfxo$. 

\begin{figure}[htb]
\begin{center}
\input{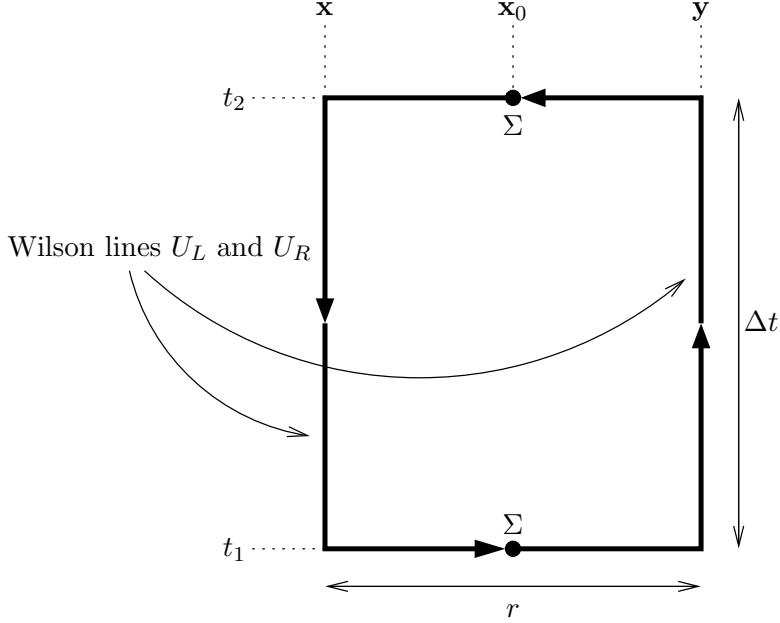}
\caption{\label{FIG005}the Wilson loop $W_\Sigma$.}
\end{center}
\end{figure}

If there exists a positive Hamiltonian, the correlation functions of these
states in Euclidean time decay exponentially with eigenvalues
of the Hamiltonian. In the static quark mass limit, $M\rightarrow\infty$, the
correlators factorise in products of free static propagators and Wilson loops,
\begin{eqnarray}
\label{EQN001} \langle \Phi^\Sigma(t_2) | \Phi^\Sigma(t_1) \rangle \ \ = \ \ e^{-2 M \Delta t}  N 
\Big\langle W_\Sigma(r,\Delta t) \Big\rangle \quad , \quad W_\Sigma \ \ \equiv 
\ \ \frac{1}{N} \textrm{Tr}\Big(\Sigma U_R \Sigma^\dagger U_L\Big)
\end{eqnarray}
with $r \equiv |\bfx-\bfy|$ and $\Delta t \equiv t_2-t_1 > 0$ (cf.\ Figure~\ref{FIG005}).
The exponential decay of the Wilson loop defines the corresponding static potentials,
\begin{eqnarray}
\Big\langle W_\Sigma(r,\Delta t) \Big\rangle \ \ = \ \ \sum_{n=0}^\infty c_n \exp\Big(-V_n^\Sigma(r) \Delta t\Big) \ \ \stackrel{\Delta t \rightarrow \infty}{\propto} \ \ \exp\Big(-V^\Sigma(r) \Delta t\Big) .
\end{eqnarray}
From the behaviour at asymptotically large time separation
the ground state potentials $V^\Sigma(r) \equiv V_0^\Sigma(r)$, 
$\Sigma \in \{ 1 , T^a \}$ can be extracted. 
This is often used to define the ground state potentials directly through the 
corresponding Wilson loops,
\begin{eqnarray}
V^\Sigma(r) \ \ = \ \ -\lim_{\Delta t \rightarrow \infty} 
\frac{\langle \dot{W}_\Sigma(r,\Delta t) \rangle}{\langle W_\Sigma(r,\Delta t) \rangle} .
\end{eqnarray}

The properties and interpretation of the potentials depend on $\Sigma$ and, in case of $\Sigma = T^a$, on the choice of the gauge fixing condition. Without gauge fixing a non-perturbative evaluation of the correlator gives
\begin{eqnarray}
\Big\langle W_{T^a}(r,\Delta t) \Big\rangle \ \ = \ \ 0 .
\end{eqnarray}
For $N = 2, 3$ the corresponding static potential is usually called triplet/octet 
static potential and $T^a = \sigma^a/2 , \lambda^a/2$.

It has been proposed to alternatively base the definition of 
the colour-adjoint static potential on a gauge invariant correlator using 
$\Sigma= T^a \mathbf{B}^a(\bfxo)$, where the adjoint transformation
behaviour of the colour-magnetic field cancels that of the strings at $\bfxo$.
This yields a non-vanishing correlation function even without gauge fixing and in leading
order of a multipole expansion receives contributions from the adjoint channel \cite{pnrqcd}.
However, we shall show in section~\ref{SEC031} that non-perturbatively
the eigenstates contributing to the exponential decay are in the colour-singlet 
sector and hence this correlator cannot be used for a non-perturbative definition of
a colour-adjoint static potential.


\section{\label{SEC632}The static potentials on the lattice}


\subsection{\label{SEC691}Temporal gauge on the lattice}

Temporal gauge $A_0^g = 0$ in the continuum corresponds to temporal links $U_0^g(t,\mathbf{x}) = 1$ on a lattice. These links gauge transform according to
\begin{eqnarray}
\ \ U_0^g(t,\mathbf{x}) \ \ = \ \ g(t,\mathbf{x}) U_0(t,\mathbf{x}) g^\dagger(t+a,\mathbf{x}) ,
\end{eqnarray}
where $g(t,\mathbf{x}) \in \textrm{SU(}N\textrm{)}$. On a lattice with finite 
temporal extent $T$ with periodic boundary conditions
it is not possible to realise temporal gauge everywhere, since gauge fixed 
links must not form closed loops. There will, hence, be one slice of links 
where $U_0 \neq 1$. In the following we take $U_0^g(t=0,\mathbf{x}) \neq 1$ 
while $U_0^g(t=1 \ldots T-1,\mathbf{x}) = 1$. A possible choice for the 
gauge transformation $g(t,\mathbf{x})$ implementing temporal gauge is
\begin{eqnarray}
g(t=2a,\mathbf{x}) & = & U_0(t=a,\mathbf{x}) , \nonumber\\
g(t=3a,\mathbf{x}) & = & g(t=2a,\mathbf{x}) U_0(t=2a,\mathbf{x}) \ \ = \ \ U_0(t=a,\mathbf{x}) U_0(t=2a,\mathbf{x}) , \nonumber\\
g(t=4a,\mathbf{x}) & = & g(t=3a,\mathbf{x}) U_0(t=3a,\mathbf{x}) \ \ = \ \ U_0(t=a,\mathbf{x}) U_0(t=2a,\mathbf{x}) U_0(t=3a,\mathbf{x}) , \nonumber \\
\label{EQN752} \ldots & = & \ldots
\end{eqnarray}
Note that this does not fix the gauge completely, i.e.\ there are residual time independent
gauge transformations $g(\mathbf{x})$ which preserve the gauge condition $U_0^g(t=1 \ldots T-1,\mathbf{x}) = 1$.


\subsection{\label{SEC942}The singlet correlator}

The trial state $| \Phi^1 \rangle$ is gauge invariant and so is the 
corresponding gauge field observable, the standard Wilson loop 
$W_1(r,\Delta t)$. Therefore, its value is identical with or without gauge fixing.
For the following, however,
it is instructive to consider the gauge dependent correlator of two spatial
strings  
\begin{eqnarray}
\Big\langle \textrm{Tr}\Big(U(t_1;\bfx,\bfy) U(t_2;\bfy,\bfx)\Big) \Big\rangle ,
\label{stringcor}
\end{eqnarray}
which vanishes without gauge fixing. On the other hand, in temporal gauge
it is non-zero and indeed 
equivalent to the manifestly gauge invariant Wilson loop, as can be seen by writing out
the gauge fixed link.
In the following we consider two cases.

\begin{figure}[htb]
\begin{center}
\input{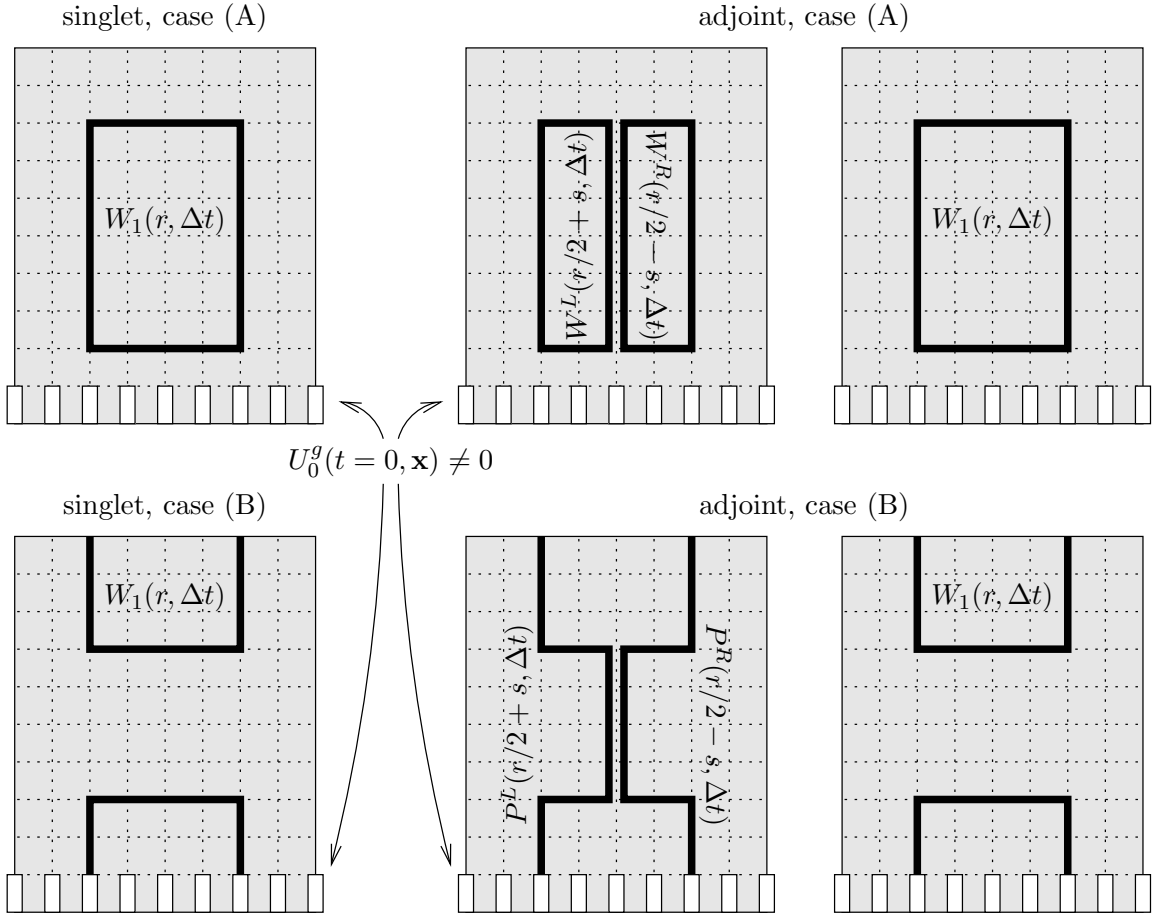}
\caption{\label{FIG003}gauge invariant observables corresponding to the gauge variant correlators defined in section~\ref{SEC942} and \ref{SEC436} in temporal gauge.}
\end{center}
\end{figure}

In case (A) the strings are correlated within the temporal lattice extent, 
i.e.~ $1 \leq t_1 < t_2 < T$, and we define $\Delta t \equiv t_2 - t_1$ (cf.\ Figure~\ref{FIG003}, left top), 
\begin{eqnarray}
\nonumber & & \hspace{-0.7cm} \Big\langle \textrm{Tr}\Big(U^g(t_1;\bfx,\bfy) U^g(t_2;\bfy,\bfx)\Big) \Big\rangle_\textrm{temporal gauge} \ \ = \\
\nonumber & & = \ \ \Big\langle \textrm{Tr}\Big(U(t_1;\bfx,\bfy) \underbrace{g^\dagger(t_1,\bfy) g(t_2,\bfy)}_{= U(t_1,t_2;\bfy)} U(t_2;\bfy,\bfx) \underbrace{g^\dagger(t_2,\bfx) g(t_1,\bfx)}_{= U(t_2,t_1;\bfx)}\Big) \Big\rangle \ \ = \\
\label{EQN861} & & = \ \ N \Big\langle W_1(r,\Delta t) \Big\rangle .
\end{eqnarray}

In case (B) the correlation is through the periodic temporal boundary, 
$0 = t_1 < t_2 < T$ (where we define $\Delta t \equiv t_2 - t_1$) or 
$1 \leq t_2 < t_1 < T$ (where we define $\Delta t \equiv t_2 - (t_1-T)$), cf.\ Figure~\ref{FIG003} (left bottom). 
The correlator now includes the unfixed temporal links and in
temporal gauge is again equivalent to the gauge invariant Wilson loop, which in this case closes
through the boundary,
\begin{eqnarray}
\nonumber & & \hspace{-0.7cm} \Big\langle \textrm{Tr}\Big(U^g(t_1;\bfx,\bfy) U^g_0(t=0,\bfy) U^g(t_2;\bfy,\bfx) (U_0^g)^\dagger(t=0,\bfx)\Big) \Big\rangle_\textrm{temporal gauge} \ \ = \\
 & & = \ \ N \Big\langle W_1^b(r,\Delta t) \Big\rangle .
\end{eqnarray}

Thus, temporal gauge turns a gauge dependent observable
into a gauge invariant observable by implicitly saturating all colour charges with 
static sources, thus ensuring gauge covariant time evolution of the charges.
The same observation will in the following be helpful
to interpret the colour-adjoint static potential.


\subsection{\label{SEC436}The adjoint correlator}

Again we distinguish the cases (A) and (B), which this time yield different results.


\subsubsection*{Case (A): $1 \leq t_1 < t_2 < T$}

In temporal gauge the correlator of two spatial strings with additional 
adjoint transformation behaviour at $\bfxo$ 
is equivalent to our observable of interest, the Wilson loop $\langle W_{T^a}(r,\Delta t) \rangle$. 
Moreover, in temporal gauge both are equivalent to a manifestly gauge invariant observable, 
as we now show. In the following
set of equations, the repeated colour index $a$ is {\it not} summed over,
\begin{eqnarray}
\nonumber & & \hspace{-0.7cm} \Big\langle \textrm{Tr}\Big(U^{T^a,g}(t_1;\bfx,\bfy) U^{(T^a)^\dagger,g}(t_2;\bfy,\bfx)\Big) \Big\rangle_\textrm{temporal gauge} \ \ = \ \ N \Big\langle W_{T^a}(r,\Delta t) \Big\rangle_\textrm{temporal gauge} \ \ = \\
\nonumber & & = \ \ \frac{1}{N^2 - 1} \sum_a \Big\langle \textrm{Tr}\Big(U(t_1;\bfx,\bfxo) g^\dagger(t_1,\bfxo) T^a g(t_1,\bfxo) U(t_1;\bfxo,\bfy) \underbrace{g^\dagger(t_1,\bfy) g(t_2,\bfy)}_{= U(t_1,t_2;\bfy)} \\
\nonumber & & \hspace{0.675cm} U(t_2;\bfy,\bfxo) g^\dagger(t_2,\bfxo) \underbrace{(T^a)^\dagger}_{= T^a} g(t_2,\bfxo) U(t_2;\bfxo,\bfy) \underbrace{g^\dagger(t_2,\bfx) g(t_1,\bfx)}_{= U(t_2,t_1;\bfx)}\Big) \Big\rangle \ \ = \\
\nonumber & & = \ \ \frac{1}{2 (N^2 - 1)} \Big\langle \underbrace{\textrm{Tr}\Big(U(t_1;\bfx,\bfxo) \underbrace{g^\dagger(t_1,\bfxo) g(t_2,\bfxo)}_{= U(t_1,t_2;\bfxo)} U(t_2;\bfxo,\bfx) U(t_2,t_1;\bfx)\Big)}_{\equiv N W^L(|\bfx-\bfxo|,\Delta t)} \\
\nonumber & & \hspace{0.675cm} \underbrace{\textrm{Tr}\Big(U(t_1;\bfxo,\bfy) U(t_1,t_2;\bfy) U(t_2;\bfy,\bfxo) \underbrace{g^\dagger(t_2,\bfxo) g(t_1,\bfxo)}_{= U(t_2,t_1;\bfxo)}\Big)}_{\equiv N W^R(|\bfy-\bfxo|,\Delta t)} \Big\rangle \\
\nonumber & & \hspace{0.675cm} -\frac{1}{2 N (N^2 - 1)} \Big\langle \textrm{Tr}\Big(U(t_1;\bfx,\bfxo) \underbrace{g^\dagger(t_1,\bfxo) g(t_1,\bfxo)}_{= 1} U(t_1;\bfxo,\bfy) U(t_1,t_2;\bfy) \\
\nonumber & & \hspace{0.675cm} U(t_2;\bfy,\bfxo) \underbrace{g^\dagger(t_2,\bfxo) g(t_2,\bfxo)}_{= 1} U(t_2;\bfxo,) U(t_2,t_1;\bfx)\Big) \Big\rangle \ \ = \\
\label{EQN225} & & = \ \ \frac{1}{2 (N^2 - 1)} \Big(N^2 \Big\langle W^L(|\bfx-\bfxo|,\Delta t) W^R(|\bfy-\bfxo|,\Delta t) \Big\rangle - \Big\langle W_1(r,\Delta t) \Big\rangle\Big) ,
\end{eqnarray}
where
\begin{eqnarray}
\sum_a T_{\alpha \beta}^a T_{\gamma \delta}^a \ \ = \ \ \frac{1}{2} \bigg(\delta_{\alpha \delta} \delta_{\beta \gamma} - \frac{1}{N} \delta_{\alpha \beta} \delta_{\gamma \delta}\bigg)
\end{eqnarray}
has been used. 
The resulting gauge invariant combination of loops is depicted in Figure~\ref{FIG003} (right top).
For an interpretation, it is better to rewrite it more compactly as
\begin{eqnarray}
\nonumber & & \hspace{-0.7cm} \Big\langle W_{T^a}(r,\Delta t) \Big\rangle_\textrm{temporal gauge} \ \ = \\
\label{EQN587} & & = \ \ \frac{2}{N(N^2-1)} \sum_a \sum_b \Big\langle \textrm{Tr}\Big(T^a U_R T^b U_L\Big) \textrm{Tr}\Big(T^a U(t_1,t_2;\bfxo) T^b U(t_2,t_1;\bfxo)\Big) \Big\rangle .
\end{eqnarray}
On the right side the first trace is our observable of interest, the Wilson loop with generator
matrix insertions, which now gets multiplied with the adjoint 
representation of a temporal Wilson line
at $\bfxo$ (second trace). The latter corresponds to a propagator of a static quark in the
adjoint representation. 
Hence, the gauge invariant expression on the right side is the 
correlation function of a gauge invariant three-quark state, 
one fundamental static quark at $\bfy$, one fundamental static anti-quark at $\bfx$, 
one adjoint static quark at $\bfxo$. Indeed, after defining
\begin{eqnarray}
 & & \hspace{-0.7cm} | \Phi^{Q \bar{Q} Q^\textrm{ad}} \rangle \ \ \equiv \ \ Q^{\textrm{ad},a}(\bfxo) \Big(\bar{Q}(\bfx) U^{T^a}(\bfx,\bfy) Q(\bfy)\Big) |0\rangle \\
 & & \hspace{-0.7cm} \langle \Phi^{Q \bar{Q} Q^\textrm{ad}}(t_2) | \Phi^{Q \bar{Q} Q^\textrm{ad}}(t_1) \rangle \ \ \equiv \ \ e^{-(2 M + M_\textrm{ad}) \Delta t} \frac{N(N^2 - 1)}{2} \Big\langle W_{Q \bar{Q} Q^\textrm{ad}}(r,\Delta t) \Big\rangle ,
\end{eqnarray}
it is easy to verify explicitly that
\begin{equation}
 2 N \Big\langle W_{T^a}(r,\Delta t) \Big\rangle_\textrm{temporal gauge} \ \ = \ \ \Big\langle W_{Q \bar{Q} Q^\textrm{ad}}(r,\Delta t) \Big\rangle .
\end{equation}
Consequently, the static potential $V^{T^a}(r)$ in temporal gauge 
should not be interpreted as the potential of a static quark and a static anti-quark, 
which form a colour-adjoint state, as has been suggested in the literature. 
$V^{T^a}(r)$ is rather a potential of a colour-singlet three-quark state 
(one fundamental static quark, one fundamental static anti-quark, 
one adjoint static quark). 
Note that the potential does not only depend on the $Q \bar{Q}$ separation 
$r=|\bfx-\bfy|$, but also on the position $s=|\bfx-\bfxo|/2-|\bfy-\bfxo|/2$ 
of the static 
adjoint quark $Q^\textrm{ad}$, i.e.\ $V^{T^a}(r,s)$. This should be kept in 
mind, while in the following we work with the symmetric alignment, where 
the adjoint source is in the centre of mass, $\bfxo=(\bfx+\bfy)/2$.


\subsubsection*{\label{SEC437} Case (B): $0 = t_1 < t_2 < T$ or $1 \leq t_2 < t_1 < T$}

Proceeding as in section~\ref{SEC942} for case (B) and in (\ref{EQN225}) one obtains
\begin{eqnarray}
\nonumber & & \hspace{-0.7cm} \Big\langle \textrm{Tr}\Big(U^{T^a,g}(t_1;\bfx,\bfy) U_0^g(t=0,\bfy) \tilde{U}^{(T^a)^\dagger,g}(t_2;\bfy,\bfx) 
 (U_0^g)^\dagger(t=0,\bfx)\Big) \Big\rangle_\textrm{temporal gauge} \ \ = \\
 & & \ \ = \ \ \frac{1}{2 (N^2 - 1)} \Big(N^2 \Big\langle P^L(r/2+s,\Delta t) 
P^R(r/2-s,\Delta t) \Big\rangle - \Big\langle W_1(r,\Delta t) \Big\rangle\Big) ,
\end{eqnarray}
which is shown in Figure~\ref{FIG003} (right bottom).

This correlation function is not suitable, to determine the potential of a 
static quark and anti-quark, which form a colour-adjoint state. 
It contains information about static colour-singlet quark anti-quark states 
propagating through the periodic boundary, and about gluelump states (a static adjoint quark 
and gluons forming a colour-singlet) propagating within the temporal lattice extent.
We will explicitly prove this statement using the transfer matrix formalism 
in section~\ref{SEC010}. 


\subsection{\label{SEC032}The transfer matrix formalism}

A useful theoretical tool to understand which states contribute to a 
correlation function is the transfer matrix formalism 
(cf.\ e.g.\ \cite{Philipsen:2001ip,jp}).
The transfer matrix propagates states by one lattice unit in time. 
In temporal gauge the transfer matrix
$\hat{T}_0$ is defined via
\begin{eqnarray}
\langle U_j(t+1) | \hat{T}_0 | U_j(t) \rangle \ \ \equiv \ \ T_{0,t+1 \leftrightarrow t} \quad , \quad T_{0,t+1 \leftrightarrow t} \ \ \equiv \ \ e^{-S(U_j(t+1),1,U_j(t))} ,
\end{eqnarray}
where $U_j(t)$ denotes the spatial links of timeslice $t$, $| U_j(t) \rangle$ is the analogue of a position eigenstate in quantum mechanics, i.e.\ a state with spatial links $U_j(t)$, and
\begin{eqnarray}
\nonumber & & \hspace{-0.7cm} S(U_j(t+1),U_0(t),U_j(t)) \ \ \equiv \ \ \frac{1}{2} S_1(U_j(t+1)) + S_2(U_j(t+1),U_0(t),U_j(t)) + \frac{1}{2} S_1(U_j(t)) \\
\label{EQN468} & & \\
 & & \hspace{-0.7cm} S_1(U_j(t)) \ \ \equiv \ \ -\frac{\beta}{2} \sum_{p \in P_t} \textrm{Re}\Big(\textrm{Tr}(U_p)\Big) \\
\label{EQN469} & & \hspace{-0.7cm} S_2(U_j(t+1),U_0(t),U_j(t)) \ \ \equiv \ \ -\frac{\beta}{2} \sum_{p \in P_{t+1,t}} \textrm{Re}\Big(\textrm{Tr}(U_p)\Big)
\end{eqnarray}
($P_t$ are space-like plaquettes on timeslice $t$ and $P_{t+1,t}$ are time-like plaquettes connecting timeslices $t$ and $t+1$), i.e.\ $S(U_j(t+1),U_0(t),U_j(t))$ is that part of the lattice Yang-Mills action containing and connecting timeslices $t$ and $t+1$ (cf.\ e.g.\ \cite{Montvay:1994cy}).

The transfer matrix $\hat{T}_0$ acts on the Hilbert space of square integrable
wave functions, including ones transforming non-trivially under the residual time independent gauge transformations $g(\mathbf{x})$.
The Hilbert space splits into orthogonal sectors with charges in arbitrary 
representations at any lattice point. Each charge sector can be isolated by appropriate
projection operators. Let $\hat{R}[g]$ be an operator to impose a gauge transformation
with gauge function $g(\bfx)$,
\begin{eqnarray}
\hat{R}[g] |\psi[U]\rangle \ \ = \ \ |\psi^g[U]\rangle \ \ = \ \ |\psi[U^g]\rangle .
\end{eqnarray}
The transfer matrix in temporal gauge commutes with time-independent 
gauge transformations, $[\hat{T}_0 , \hat{R}[g]] = 0$, which implies
gauge invariance of its eigenvalues.
Another consequence is that eigenstates of $\hat{T}_0$ can simultaneously be chosen as eigenstates of 
$\hat{R}[g]$, i.e.\ they can be classified according to irreducible 
$\textrm{SU(}N\textrm{)}$ colour multiplets at each $\mathbf{x}$. 
E.g.\ for the gauge group $SU(2)$ these have the same structure as spin/angular 
momentum multiplets, which are well-known from ordinary quantum mechanics.
 
Specifically, we list the transformation behaviour of states occurring in our 
analysis: $i)$ colour-singlets, $ii)$   
states with a fundamental charge at $\bfy$ and an anti-fundamental one
at $\bfx$, $iii)$ states with an adjoint charge at $\bfxo$ and finally $iv)$ states
with a fundamental charge at $\bfy$, an anti-fundamental one
at $\bfx$ and an adjoint charge at $\bfxo$,
\begin{eqnarray}
\label{trafo}
i) & & \hat{R}[g]|\psi\rangle \ \ = \ \ |\psi\rangle \\
\label{trafof}
ii) & & \hat{R}[g] |\psi_{\alpha\beta}\rangle
\ \ = \ \ g_{\alpha\gamma}(\bfx) 
g^\dagger_{\delta\beta}(\bfy)|\psi_{\gamma\delta}\rangle \\
\label{trafogl}
iii)& & 
 \hat{R}[g] |\psi^a\rangle \ \ = \ \ D^A_{ab}(g(\bfxo))|\psi^b\rangle\\
\label{trafoa}
iv) & &\hat R[g] |\psi^a_{\alpha\beta}\rangle
\ \ = \ \ g_{\alpha\gamma}(\bfx)
g^\dagger_{\delta\beta}(\bfy)D^{\rm A}_{ab}(g(\bfx_0))|\psi^b_{\gamma\delta}\rangle  ,
\end{eqnarray}
where $D^{\rm A}_{a b}(g)$ are
the representation matrices of the adjoint representation, 
\begin{eqnarray}
D^{\rm A}_{a b}(g) \ \ = \ \ 2\,\tr(g^\dag T^agT^b).  
\end{eqnarray}
The projectors onto the corresponding orthogonal charge sectors of the Hilbert space 
are
\begin{eqnarray}
i) & &\hat P \ \ = \int D g\, \hat R[g]\\
ii) & & \hat P^{{\rm F}\otimes\bar{\rm F}}_{\alpha\beta\mu\nu} 
\ \ = \ \ \int D g\, g^\dagger_{\alpha\beta}(\bfx)\, g_{\mu\nu}(\bfy) \,\hat R[g]\\
iii)& & \hat{P}^A_{ab} \ \ = \ \ \int Dg\,D^A_{ab}(g^\dagger(\bfxo))\hat{R}[g]\\ 
iv) & &\hat{P}_{\alpha\beta\mu\nu a b}^{{\rm F}\otimes\bar{\rm F}\otimes {\rm A}} \ \ = \ \ \int
Dg \, g^\dag_{\alpha\beta}(\bfx)g_{\mu\nu}(\bfy)
D^{\rm A}_{ab}(g^\dagger(\bfx_0))\hat{R}[g] .
\end{eqnarray}
For example, $ii)$ maps the component $|\psi_{\beta\mu}\rangle$ of a
representation ${\rm F}\otimes\bar{\rm F}$ to the component $|\psi_{\alpha\nu}\rangle$
and annihilates all other components and charge sectors\footnote{This is 
due to the fact that only group integrals over the trivial representation 
are non-zero
\cite{Creutz:1978ub},
$\int Dg =1$, $\int Dg \, g_{\alpha \beta} = 0$, $\int dg \, g_{i j} g_{k l}^\dagger=(1/N) \delta_{i l} \delta_{j k}$, ...}.
Direct calculation shows
\begin{eqnarray}
\hat P^{{\rm F}\otimes\bar{\rm F}}_{\alpha\beta\mu\nu} |\psi_{\gamma\delta}\rangle
 \ \ = \ \ \frac{1}{N^2} \delta_{\beta\gamma} \delta_{\mu\delta} |\psi_{\alpha\nu}\rangle \quad , \quad
\hat P^{{\rm F}\otimes\bar{\rm F}}_{\alpha\beta\mu\nu} |\psi_{\beta\mu}\rangle
 \ \ = \ \ |\psi_{\alpha\nu}\rangle .
\end{eqnarray}
Similarly, one verifies
\begin{eqnarray}
\hat P^{{\rm F}\otimes\bar{\rm F}\otimes A}_{\alpha\beta\mu\nu ab}
|\psi^c_{\gamma\delta}\rangle
\ \ = \ \ \frac{1}{N^2(N^2-1)} \delta_{\beta\gamma} \delta_{\mu\delta}\delta_{bc}
|\psi^a_{\alpha\nu}\rangle \quad , \quad
\hat P^{{\rm F}\otimes\bar{\rm F}\otimes A}_{\alpha\beta\mu\nu ab} 
|\psi^b_{\beta\mu}\rangle
\ \ = \ \ |\psi^a_{\alpha\nu}\rangle .
\end{eqnarray}
%


\subsection{\label{SEC031}Spectral analysis of Wilson loops, case (A)}

We are now ready to perform the spectral analysis of our lattice observables in temporal gauge. We start by considering those observables, where the corresponding Wilson loops do not include links $U_0(t=0,\mathbf{x}) \neq 1$ (cf.\ Figure~\ref{FIG003}, ``singlet, case (A)'' and ``adjoint, case (A)''). To this end we write out the Euclidean path integral in terms of the transfer matrix, 
\begin{eqnarray}
\nonumber & & \hspace{-0.7cm} N \Big\langle W_\Sigma(r,\Delta t) \Big\rangle \ \ = \ \ \frac{1}{Z} \int DU_j \, \int DU_0(0) \, U^\Sigma_{\alpha \beta}(t_1;\bfx,\bfy) U^{\Sigma^\dagger}_{\beta \alpha}(t_2;\bfy,\bfx) \\
\label{EQN202} & & \hspace{0.675cm} T_{0,0 \leftrightarrow T-1} \ldots T_{0,3 \leftrightarrow 2} T_{0,2 \leftrightarrow 1} e^{-S(U_j(1),U_0(0),U_j(0))} ,
\end{eqnarray}
where $\int DU_j$ denotes the integration over all spatial links and 
$\int DU_0(0)$ the integration over all temporal links connecting 
timeslice $t=0$ and $t=1$.
The path integral 
can now be rewritten according to
\begin{eqnarray}
\nonumber & & \hspace{-0.7cm} N \Big\langle W_\Sigma(r,\Delta t) \Big\rangle \ \ = \\
\nonumber & & = \ \ \frac{1}{Z} \int DU_j \, \int DU_0(0) \, U^\Sigma_{\alpha \beta}(t_1;\bfx,\bfy) U^{\Sigma^\dagger}_{\beta \alpha}(t_2;\bfy,\bfx) \\
\nonumber & & \hspace{0.675cm} T_{0,0 \leftrightarrow T-1} \ldots T_{0,2 \leftrightarrow 1} e^{-S(U_j(1),U_0(0),U_j(0))} \ \ = \\
\nonumber & & = \ \ \frac{1}{Z} \int DU_j \, \int Dg \, U^\Sigma_{\alpha \beta}(t_1;\bfx,\bfy) U^{\Sigma^\dagger}_{\beta \alpha}(t_2;\bfy,\bfx) T_{0,0 \leftrightarrow T-1} \ldots T_{0,2 \leftrightarrow 1} e^{-S(U_j(1),1,U_j^{g^\dagger}(0))} \ \ = \\
\nonumber & & = \ \ \frac{1}{Z} \textrm{Tr}\bigg(\hat{T}_0^{T - t_2} \hat{U}^{\Sigma^\dagger}_{\beta \alpha}(\bfy,\bfx) \hat{T}_0^{t_2 - t_1} \hat{U}^{\Sigma}_{\alpha \beta}(\bfx,\bfy) \hat{T}_0^{t_1} \hat{P}\bigg) \ \ = \\
\nonumber & & = \ \ \frac{1}{Z} \sum_{k,m,n} \langle n | \hat{T}_0^{T - t_2} \hat{U}^{\Sigma^\dagger}_{\beta \alpha}(\bfy,\bfx) | k \rangle \langle k | \hat{T}_0^{t_2 - t_1} \hat{U}^\Sigma_{\alpha \beta}(\bfx,\bfy) | m \rangle \langle m | \hat{T}_0^{t_1} \hat{P} | n \rangle \ \ = \\
\nonumber & & = \ \ \frac{1}{Z} \sum_{k,m,n} e^{-E_n (T - t_2)} e^{-E_k (t_2 - t_1)} e^{-E_m t_1} \langle n | (\hat{U}^{\Sigma}_{\alpha \beta}(\bfx,\bfy))^\dagger | k \rangle \langle k | \hat{U}^\Sigma_{\alpha \beta}(\bfx,\bfy) | m \rangle \langle m |\hat{P} | n \rangle . \\
 & &
\end{eqnarray}
$\hat{P} | n \rangle \neq 0$ only, if $| n \rangle$ is a gauge invariant state, i.e.\ if $| n \rangle$ is a colour-singlet. Then
\\ $\langle m | \hat{P} | n \rangle = \delta_{m n}$
and 
\begin{eqnarray}
 N  \Big\langle W_\Sigma(r,\Delta t) \Big\rangle \ \ = \ \ \frac{1}{Z} \sum_{k,n'} e^{-E_k \Delta t} e^{-E_{n'} (T - \Delta t)} \sum_{\alpha,\beta} \Big|\langle k | \hat{U}^\Sigma_{\alpha \beta}(\bfx,\bfy) | n' \rangle\Big|^2 ,
\end{eqnarray}
where $\sum_{n'}$ is over gauge invariant states $| n' \rangle$ only. 
The nature of the states $|k\rangle$ is determined by the choice of $\Sigma$.

For $\Sigma=1$ the state $\hat{U}^\Sigma_{\alpha \beta}(\bfx,\bfy) | n' \rangle = \hat{U}_{\alpha \beta}(\bfx,\bfy) | n' \rangle$ 
transforms under gauge transformations according to (\ref{trafof}).
Hence, 
\begin{eqnarray}
\langle k|\hat{U}_{\alpha \beta}(\bfx,\bfy) | n' \rangle \ \ = \ \ 
\langle k|\hat{P}^{{\rm F}\otimes\bar{\rm F}}_{\alpha\mu\nu\beta}
\hat{U}_{\mu \nu}(\bfx,\bfy) | n' \rangle
\end{eqnarray}
such that only states $\langle k'_{\alpha\beta}|$ with the same transformation 
behaviour contribute to the sum, while all others are annihilated, 
\begin{eqnarray}
\label{EQN210} N \Big\langle W_1(r,\Delta t) \Big\rangle \ \ = \ \ 
\frac{1}{Z} \sum_{k',n'} e^{-V_{k'}^1(r) \Delta t} 
e^{-\mathcal{E}_{n'} (T - \Delta t)} \sum_{\alpha,\beta} 
\Big|\langle k'_{\alpha \beta} | \hat{U}_{\alpha \beta}(\bfx,\bfy) | n' \rangle\Big|^2 
\end{eqnarray}
($\mathcal{E}_{n'}$ denotes eigenvalues of gauge invariant eigenstates of $\hat{T}_0$, 
i.e.\ states without static quarks).
Using the spectral decomposition of the partition function,
\begin{eqnarray}
\label{EQN208} Z \ \  = \ \ \sum_{k'} e^{-\mathcal{E}_{k'} T},
\end{eqnarray}
and considering infinite temporal lattice extent reduces $\sum_{n'}$ to the vacuum,
\begin{eqnarray}
\nonumber & & \hspace{-0.7cm}  N  \Big\langle W_1(r,\Delta t) \Big\rangle \ \ = \ \ \frac{\sum_{k',n'} e^{-V_{k'}^1(r) \Delta t} e^{-\mathcal{E}_{n'} (T - \Delta t)} \sum_{\alpha,\beta} \Big|\langle k'_{\alpha \beta} | \hat{U}_{\alpha \beta}(\bfx,\bfy) | n' \rangle\Big|^2}{\sum_{k'} e^{-\mathcal{E}_{k'} T}} \ \ \stackrel{T\rightarrow \infty}{=} \\
\label{EQN734} & & \stackrel{T\rightarrow \infty}{=} \ \ \sum_{k'} e^{-(V_{k'}^1(r) - \mathcal{E}_0) \Delta t} \sum_{\alpha,\beta} \Big|\langle k'_{\alpha \beta} | \hat{U}_{\alpha \beta}(\bfx,\bfy) | 0 \rangle\Big|^2 .
\end{eqnarray}
This is of course the well-known result for the colour-singlet static potential.
Note that there is a sum over $\alpha$ and $\beta$ in (\ref{EQN210}) and (\ref{EQN734}), which is a sum over the $N^2$ possible colour orientations of the static quark and the static anti-quark. Both the eigenvalues of the transfer matrix $V_{k'}^1(r)$ and the corresponding matrix elements $|\langle k'_{\alpha \beta} | \hat{U}_{\alpha \beta}(\bfx,\bfy) | n' \rangle|^2$ (no sum over $\alpha$ and $\beta$) are independent of $\alpha$ and $\beta$, i.e.\ each quark anti-quark colour orientation yields the same contribution to the correlation function. In other words in temporal gauge the colour orientations of the static quark and the static anti-quark are irrelevant, i.e.\ any colour orientation will result in the singlet static potential.

The result for the colour-adjoint case, i.e.\ $\Sigma=T^a$, follows in complete analogy. 
In this case the state $\hat{U}^\Sigma_{\alpha \beta}(\bfx,\bfy) | n' \rangle = \hat{U}^{T^a}_{\alpha \beta}(\bfx,\bfy) | n' \rangle$ transforms as in (\ref{trafoa}) and 
we have
\begin{eqnarray}
\langle k|\hat{U}^{T^a}_{\alpha \beta}(\bfx,\bfy) | n' \rangle \ \ = \ \ 
\langle k|\hat{P}^{{\rm F}\otimes\bar{\rm F}\otimes A}_{\alpha\mu\nu\beta ab}
\,\hat{U}^b_{\mu \nu}(\bfx,\bfy) | n' \rangle
\end{eqnarray}
such that only states $\langle k'^a_{\alpha \beta} |$ with the same transformation behaviour contribute to the sum. The final result is
\begin{eqnarray}
 N \Big\langle W_{T^a}(r,\Delta t) \Big\rangle \ \ \stackrel{T \rightarrow \infty}{=} \ \ \sum_{k'} e^{-(V_{k'}^{T^a}(r) - \mathcal{E}_0) \Delta t}
\sum_{\alpha,\beta} \Big|\langle k'^a_{\alpha \beta} | \hat{U}^a_{\alpha \beta}(\bfx,\bfy)
|0\rangle\Big|^2.
\end{eqnarray}
This correlation function is suited to extract a three-quark potential of a 
fundamental static quark, a fundamental static anti-quark and an adjoint static quark. 
It is not a quark anti-quark potential with the quark and the anti-quark in a 
colour-adjoint orientation, i.e.\ it should not be interpreted as a colour-adjoint 
static potential. 
As for $\Sigma = 1$ the potential is 
independent of the colour orientations of the three quarks.
Cf.\ also section~\ref{SEC436}, where the same result has been obtained using different methods and arguments.

Finally, in the case $\Sigma = T^a \mathbf{B}^a$ the state 
$\hat{U}^\Sigma_{\alpha \beta}(\bfx,\bfy) | n' \rangle = \hat{U}^{T^a \mathbf{B}^a}_{\alpha\beta}(\bfx,\bfy) |n'\rangle$
again transforms as in (\ref{trafof}), i.e.\ has fundamental charges at $\bfx$ and $\bfy$. Hence,
\begin{eqnarray}
\langle k|\hat{U}^{T^a \mathbf{B}^a}_{\alpha \beta}(\bfx,\bfy) | n' \rangle \ \ = \ \ 
\langle k|\hat{P}^{{\rm F}\otimes\bar{\rm F}}_{\alpha\mu\nu\beta}
\hat{U}^{T^a \mathbf{B}^a}_{\mu \nu}(\bfx,\bfy) | n' \rangle
\end{eqnarray}
and thus only states $\langle k'_{\alpha\beta}|$ contribute. These are in the same colour charge
sector as those in (\ref{EQN210}), however, in this case their parity is negative (for a more detailed discussion regarding quantum numbers [other than colour] of states with two static charges we refer to e.g.\ \cite{Bali:2005fu,Wagner:2010ad}).
The final result for this case then reads
\be
\Big\langle W_{B}(r,\Delta t) \Big\rangle \ \ \stackrel{T\rightarrow \infty}{\longrightarrow} \ \
\sum_{k'} e^{-(V_{k'}^{1,-}(r) - \mathcal{E}_0) \Delta t}
\sum_{\alpha,\beta} \Big|\langle k'_{\alpha \beta} | \hat{U}^{T^a \mathbf{B}^a}_{\alpha \beta}(\bfx;\bfy)
|0\rangle\Big|^2   . 
\ee   
The exponential decay is with colour-singlet potentials in the negative parity channel, 
thus $\langle W_B(r,\Delta t) \rangle$ 
is not suitable, to define a colour-adjoint static potential.


\subsection{\label{SEC010} Spectral analysis of Wilson loops, case (B)}

In this section we give the spectral decomposition for the situation, where the correlators close 
through the temporal boundary. 
The backwards Wilson loop, cf.\ Figure~\ref{FIG003} (left bottom) can be rewritten according to
\begin{eqnarray}
\nonumber & & \hspace{-0.7cm} N \Big\langle W_1^b(r,\Delta t) \Big\rangle \ \ = \\
\nonumber & & = \ \ \frac{1}{Z} \int DU_j \, \int DU_0(0) \, U_{0,\alpha\beta}^\dagger(t=0,\bfx) U_{\beta \gamma}(t_1;\bfx,\bfy) U_{0,\gamma\delta}(t=0,\bfy) U_{\delta \alpha}(t_2;\bfy,\bfx) \\
\nonumber & & \hspace{0.675cm} T_{0,0 \leftrightarrow T-1} \ldots T_{0,3 \leftrightarrow 2} T_{0,2 \leftrightarrow 1} e^{-S(U_j(1),U_0(0),U_j(0))} \ \ = \\
\nonumber & & = \ \ \frac{1}{Z} \int DU_j \, \int Dg \, g^\dagger_{\alpha \beta}(\bfx) U_{\beta \gamma}(t_1;\bfx,\bfy) g_{\gamma \delta}(\bfy) U_{\delta \alpha}(t_2;\bfy,\bfx) \\
\nonumber & & \hspace{0.675cm} T_{0,0 \leftrightarrow T-1} \ldots T_{0,3 \leftrightarrow 2} T_{0,2 \leftrightarrow 1} e^{-S(U_j(1),1,U_j^{g^\dagger}(0))} \ \ = \\
\nonumber & & = \ \ \frac{1}{Z} \sum_{k,m,n} e^{-E_n (T - t_1)} e^{-E_k (t_1 - t_2)} e^{-E_m t_2} \langle n | \hat{U}_{\beta \gamma}(\bfx,\bfy) | k \rangle \langle k | (\hat{U}_{\alpha \delta}^\dagger(\bfx,\bfy)) | m \rangle \langle m | \hat{P}^{{\rm F}\otimes\bar{\rm F}}_{\alpha\beta\gamma\delta} | n \rangle . \\
\label{EQN220} & &
\end{eqnarray}
Because of the projector, $\sum_{m,n}$ can be restricted to all eigenstates 
which transform as in (\ref{trafof}) and $\langle m'|n'\rangle=\delta_{m'n'}$.
The other matrix elements then require
$\hat{U}_{\beta\gamma}(\bfx,\bfy)|k\rangle$ to transform in the same way,
which implies $|k\rangle$ to be in the singlet sector. We finally obtain
\begin{eqnarray}
 N \Big\langle W_1^b(r,\Delta t) \Big\rangle \ \ = \ \ \frac{1}{Z} \sum_{k',n'} e^{-V_{n'}^1 \Delta t} e^{-\mathcal{E}_{k'} (T - \Delta t)} \sum_{\beta \gamma} \Big|\langle n'_{\beta \gamma} | \hat{U}_{\beta \gamma}(\bfx,\bfy) | k' \rangle\Big|^2 .
\end{eqnarray}
This result is identical to (\ref{EQN210}), i.e.\ 
as expected the position of the Wilson loop relative to the slice 
of links $U_0(t=0,\mathbf{x}) \neq 1$ does not matter.

Similarly one can study the backwards correlator of the adjoint 
string $U^{T^a}_{\alpha \beta}(\bfx,\bfy)$, cf.\ Figure~\ref{FIG003} (right bottom). In this case  
\begin{eqnarray}
\nonumber & & \hspace{-0.7cm}  N \Big\langle W_{T^a}^b(r,\Delta t) \Big\rangle \ \ = \\
\nonumber & & = \ \ \frac{1}{Z} \int DU_j \, \int DU_0(0) \, U_{0,\alpha\beta}^\dagger(t=0,\bfx) U^{T^a}_{\beta \gamma}(t_1;\bfx,\bfy) U_{0,\gamma\delta}(t=0,\bfy) U^{(T^a)^\dagger}_{\delta \alpha}(t_2;\bfy,\bfx) \\
\nonumber & & \hspace{0.675cm} T_{0,0 \leftrightarrow T-1} \ldots T_{0,3 \leftrightarrow 2} T_{0,2 \leftrightarrow 1} e^{-S(U_j(1),U_0(0),U_j(0))} \ \ = \\
\nonumber & & = \ \ \frac{1}{Z} \sum_{k,m,n} e^{-E_n (T - t_1)} e^{-E_k (t_1 - t_2)} e^{-E_m t_2} \\
\nonumber & & \hspace{0.675cm} \langle n | \hat{U}^{T^a}_{\beta \gamma}(\bfx,\bfy) | k \rangle \langle k | (\hat{U}^{T^a}_{\alpha \delta}(\bfx,\bfy))^\dagger | m \rangle \langle m | \hat{P}^{{\rm F}\otimes\bar{\rm F}}_{\alpha\beta\gamma\delta} | n \rangle \ \ = \\
\label{EQN221} & & = \ \ \frac{1}{N^2 Z} \sum_{k',n'} e^{-V_{n'}^1(r) \Delta t} e^{-\mathcal{E}_{k'}^{Q^\textrm{adj}} T} \Big|\langle n'_{\beta \gamma} | \hat{U}^{T^a}_{\beta \gamma}(\bfx,\bfy) | k'^a \rangle\Big|^2 ,
\end{eqnarray}
where we have again used that the projector restricts $|m\rangle$ and $|n\rangle$ to states transforming as in (\ref{trafof}) and
$m'=n'$. Now $(\hat{U}^{T^a}_{\alpha \delta}(\bfx,\bfy))^\dagger | m'_{\alpha \delta} \rangle$ transforms as in (\ref{trafogl}), i.e.\ the states $| k \rangle$ are required to transform as a single adjoint charge at $\bfxo$ (the corresponding eigenvalues of the transfer matrix are denoted by $\mathcal{E}_{k'}^{Q^\textrm{adj}}$).

In the limit $T \rightarrow \infty$ this simplifies to
\begin{eqnarray}
 N \Big\langle W_{T^a}^b(r,\Delta t) \Big\rangle \ \ = \ \ \sum_{k',n'} e^{-(V_{n'}^1(r) - \mathcal{E}_0^{Q^\textrm{adj}}) \Delta t} e^{-(\mathcal{E}_0^{Q^\textrm{adj}} - \mathcal{E}_0) T} 
\sum_{\beta ,\gamma} \Big|\langle n'_{\beta \gamma} | 
\hat{U}^{T^a}_{\beta \gamma}(\bfx,\bfy) | k'^a \rangle\Big|^2 .
\end{eqnarray}
This correlation function is suited to extract the common singlet potential $V^1(r) = V_0^1(r)$ and a gluelump mass $\mathcal{E}_0^{Q^\textrm{adj}}$. Cf.\ also section~\ref{SEC437}, where the same result has been obtained using different methods and arguments.

\subsection{\label{sec:coul} Coulomb gauge on the lattice}

Another gauge featuring a transfer matrix/Hamiltonian is Coulomb 
gauge \cite{Zwanziger:1995cv}.
Similar to the analysis presented in section~\ref{SEC436} one can write the 
colour-adjoint Wilson loop as
\begin{eqnarray}
\nonumber & & \hspace{-0.7cm} \Big\langle W_{T^a}(r,\Delta t) \Big\rangle_\textrm{Coulomb gauge} \ \ = \ \ \frac{1}{N} \Big\langle \textrm{Tr}\Big(T^a U_R T^a U_L\Big) \Big\rangle_\textrm{Coulomb gauge} \ \ = \\
\nonumber & & = \ \ \frac{1}{N} \Big\langle \textrm{Tr}\Big(T^a g(t_1,\mathbf{x}_0) U_R g^\dagger(t_2,\mathbf{x}_0) T^a g(t_2,\mathbf{x}_0) U_L g^\dagger(t_1,\mathbf{x}_0)\Big) \Big\rangle \ \ = \\
\nonumber & & = \ \ \frac{1}{N} \Big\langle \textrm{Tr}\Big(T^a h^\textrm{res}(t_1) g^\textrm{Coulomb}(t_1,\mathbf{x}_0) U_R g^{\textrm{Coulomb},\dagger}(t_2,\mathbf{x}_0) h^{\textrm{res},\dagger}(t_2) \\
 & & \hspace{0.675cm} T^a h^\textrm{res}(t_2) g^\textrm{Coulomb}(t_2,\mathbf{x}_0) U_L g^{\textrm{Coulomb},\dagger}(t_1,\mathbf{x}_0) h^{\textrm{res},\dagger}(t_1)\Big) \Big\rangle .
\end{eqnarray}
Since the Coulomb gauge condition does not fix the gauge completely 
(a space-independent residual gauge symmetry remains), we have split 
the gauge transformation according to $g(t,\mathbf{x}) \equiv h^\textrm{res}(t) g^\textrm{Coulomb}(t,\mathbf{x})$, where $g(t,\mathbf{x})^\textrm{Coulomb} \in SU(N)$ transforms the links to Coulomb gauge and $h^\textrm{res}(t) \in SU(N)$ represents the residual gauge degrees of freedom.

Without imposing any additional gauge condition restricting $h^\textrm{res}(t)$ one obtains in a non-perturbative evaluation
\begin{eqnarray}
\Big\langle W_{T^a}(r,\Delta t) \Big\rangle_\textrm{Coulomb gauge} \ \ = \ \ 0 .
\end{eqnarray}
Because of the integration over $h^\textrm{res}(t)$ the colour-adjoint Wilson loop averages to zero (note that again the parallel transport in time is missing). 

One can complete the gauge fixing by e.g.\ also requiring 
$U_0^g(t,\mathbf{z_0}) = 1$, where $\mathbf{z_0}$ is an arbitrary 
point in space. This condition imposes temporal gauge 
(cf.\ section~\ref{SEC691}) to temporal 
links at $\mathbf{z_0}$. $h^\textrm{res}(t)$ is then chosen as specified by 
(\ref{EQN752}), replacing $g$ by $h^\textrm{res}$ and $\mathbf{x}$ by $\mathbf{z}_0$. The Wilson loop average is then non-vanishing. Similar to (\ref{EQN587}) it can be written as
\begin{eqnarray}
\nonumber & & \hspace{-0.7cm} \Big\langle W_{T^a}(r,\Delta t) \Big\rangle_\textrm{Coulomb gauge} \ \ = \\
\nonumber & & = \ \ \frac{2}{N (N^2-1)} \sum_a \sum_b \\
\nonumber & & \hspace{0.675cm} \Big\langle \textrm{Tr}\Big(T^a g^\textrm{Coulomb}(t_1,\mathbf{x}_0) U_R g^{\textrm{Coulomb},\dagger}(t_2,\mathbf{x}_0) T^b g^\textrm{Coulomb}(t_2,\mathbf{x}_0) U_L g^{\textrm{Coulomb},\dagger}(t_1,\mathbf{x}_0)\Big) \\
 & & \hspace{0.675cm} \textrm{Tr}\Big(T^a U(t_1,t_2;\bfzo) T^b U(t_2,t_1;\bfzo)\Big) \Big\rangle .
\end{eqnarray}
The second trace, which arises due to $h^\textrm{res}$, 
corresponds to a propagator of a static quark in the adjoint representation 
at $\mathbf{z}_0$. Consequently the colour-adjoint static potential 
in this ``completed Coulomb gauge'' should also be interpreted as a potential of 
three static quarks, one fundamental static quark at $\bfy$, 
one fundamental static anti-quark at $\bfx$ and one adjoint static quark 
at $\bfzo$. For the choice $\bfzo=\bfxo$ the resulting potential
agrees with that in temporal gauge (with modified matrix elements).
This is clearly not in line with the perturbative 
result (\ref{EQN110}), where no adjoint static quark exists. 
Moreover, note that the colour-adjoint static potential in completed 
Coulomb gauge depends on $\bfzo$, which is now part of the gauge condition. 
Thus, even within this class of gauges the colour-adjoint static 
potential is gauge dependent and depends on the details of the gauge condition.


\section{\label{SEC118}Leading order perturbative calculations}

While some of the following calculations are neither new nor original, 
we review the leading order perturbative results for the correlators 
discussed here in Lorenz as well as in Coulomb gauge. 


\subsection{\label{SEC119}Lorenz gauge}

The gluon propagator in Lorenz gauge $\partial_\mu A_\mu = 0$ is
\begin{eqnarray}
D_{F,\mu \nu}^{a b}(x,y) \ \ = \ \ \frac{1}{(2 \pi)^4} \int d^4p \, e^{-i p (x-y)} \delta^{a b} \frac{1}{p^2} \bigg(\delta_{\mu \nu} - (1 - \xi) \frac{p_\mu p_\nu}{p^2}\bigg) .
\end{eqnarray}
In the following we use $\xi = 1$, i.e.\
\begin{eqnarray}
\label{EQN101} D_{F,\mu \nu}^{a b}(x,y) \ \ = \ \ \delta^{a b} \delta_{\mu \nu} \frac{1}{(2 \pi)^4} \int d^4p \, e^{-i p (x-y)} \frac{1}{p^2} \ \ = \ \ \delta^{a b} \delta_{\mu \nu} \frac{1}{4 \pi^2 (x-y)^2} .
\end{eqnarray}

\begin{figure}[htb]
\begin{center}
\input{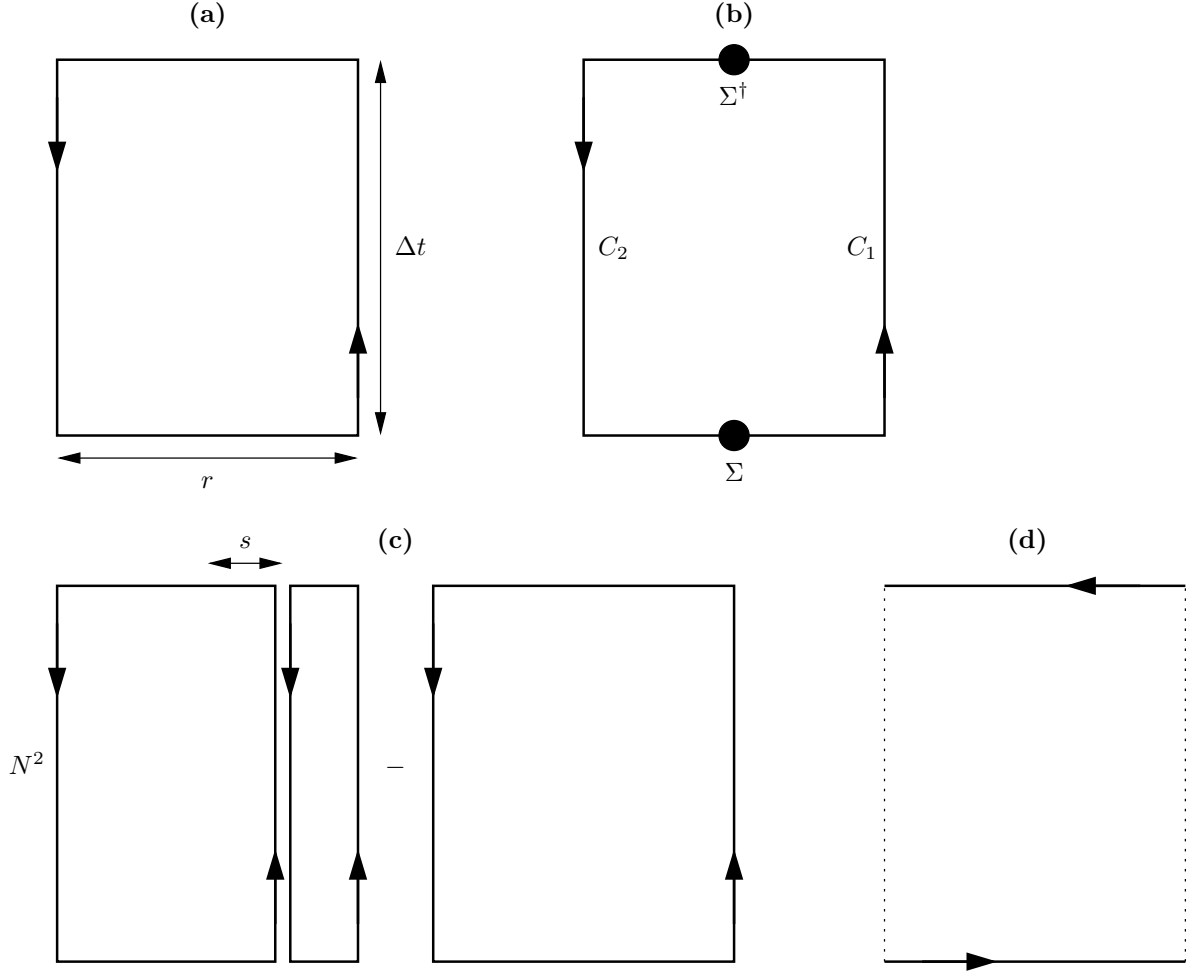}
\caption{\label{FIG002}loops and diagrams calculated perturbatively in section~\ref{SEC118}.}
\end{center}
\end{figure}

The perturbative expansion of the Wilson loop (\ref{EQN001}) with $\Sigma=1,T^a$ is
(cf.\ Figure~\ref{FIG002}(a) and (b))
\begin{eqnarray}
\nonumber & & \hspace{-0.7cm} \Big\langle W_\Sigma(r,\Delta t) \Big\rangle \ \ = \ \ \bigg\langle \frac{1}{N} \textrm{Tr}\bigg(\Sigma P \exp\bigg(i g \int_{C_1} dz_\mu \, A_\mu(z)\bigg)\Sigma^\dag P\exp\bigg(i g \int_{C_2} dz_\nu \, A_\nu(z)\bigg)\bigg)\bigg\rangle \ \ = \\
\label{pert} & & = \ \ \frac{1}{N} \textrm{Tr}(\Sigma \Sigma^\dag) - \frac{g^2}{N} 
\textrm{Tr}\Big(\Sigma T^a \Sigma^\dag T^b\Big)\int_{C_1} dx_\mu \, 
\int_{C_2} dy_\nu \, D_{F,\mu \nu}^{a b}(x,y)\nonumber \\
& &  - \ \ \frac{g^2}{2N}\textrm{Tr}\Big(\Sigma T^a T^b \Sigma^\dag \Big)\int_{C_1} dx_\mu \, 
\int_{C_1} dy_\nu \, D_{F,\mu \nu}^{a b}(x,y)\nonumber \\ 
& & - \ \ \frac{g^2}{2N}\textrm{Tr}\Big( \Sigma \Sigma^\dag T^aT^b \Big)\int_{C_2} dx_\mu \,   
\int_{C_2} dy_\nu \, D_{F,\mu \nu}^{a b}(x,y)
+ \mathcal{O}(g^4) .
\end{eqnarray}

In the limit of interest, $\Delta t\rightarrow \infty$, the contribution of
the spatial strings is suppressed and
the the integrals receive contributions
from the same temporal parallel transporter,
\begin{eqnarray}
\nonumber & & \hspace{-0.7cm} \lim_{\epsilon \rightarrow 0} 
\int_{-\Delta t/2}^{+\Delta t/2} dt_1 \, \int_{-\Delta t/2}^{+\Delta t/2} dt_2 \, \frac{1}{4 \pi^2 ((t_1 - t_2)^2 + \epsilon^2)} \ \ = \\
\nonumber & & = \ \ \frac{1}{4 \pi^2} \lim_{\epsilon \rightarrow 0} 
\bigg(+\frac{2 \Delta t}{\epsilon} \textrm{arctan}(\Delta t/\epsilon) 
- \ln\bigg(\frac{\Delta t^2}{\epsilon^2}\bigg)\bigg) \ \ = \ \ \frac{1}{4 \pi^2} \lim_{\epsilon \rightarrow 0} \bigg(+\frac{\pi \Delta t}{\epsilon} - \ln\bigg(\frac{\Delta t^2}{\epsilon^2}\bigg)\bigg) , \\
 & &
\end{eqnarray}
or from opposite temporal parallel transporters,
\begin{eqnarray}
\nonumber & & \hspace{-0.7cm} \int_{-\Delta t/2}^{+\Delta t/2} dt_1 \, \int_{+\Delta t/2}^{-\Delta t/2} dt_2 \, \frac{1}{4 \pi^2 ((t_1 - t_2)^2 + r^2)} \ \ = \\
 & & = \ \ \frac{1}{4 \pi^2} \bigg(-\frac{2 \Delta t}{r} \textrm{arctan}(\Delta t/r) + \ln\bigg(\frac{\Delta t^2 + r^2}{r^2}\bigg)\bigg) .
\end{eqnarray}
One thus obtains
\begin{eqnarray}
\label{EQN103} \lim_{\Delta t \rightarrow \infty} \Big\langle W_{\Sigma}(r,\Delta t) \Big\rangle \ \ = \ \ \frac{1}{N} \textrm{Tr}(\Sigma \Sigma^\dagger) - g^2 \bigg(V_0(\Sigma) - \frac{\textrm{Tr}(\Sigma T^a \Sigma^\dagger T^a)}{4 N \pi r}\bigg) \Delta t + \mathcal{O}(g^4) ,
\end{eqnarray}
where $V_0$ is an infinite constant, i.e.\ independent of $r$.

We begin by discussing the standard Wilson loop, $\Sigma=1$.
In this case $\textrm{Tr}(\Sigma \Sigma^\dagger) = N$ and 
$\textrm{Tr}(\Sigma T^a \Sigma^\dagger T^a) = (N^2-1) / 2$.
In Lorenz gauge a transfer matrix or Hamiltonian does not exist. 
However, for the manifestly gauge invariant Wilson loop 
the time evolution is the same in any gauge and without gauge fixing. Therefore, 
for large temporal separations $\Delta t$ they are guaranteed to decay 
exponentially proportional to the eigenvalue of the lowest energy eigenstate with corresponding 
quantum numbers,
\begin{eqnarray}
\nonumber & & \hspace{-0.7cm} \lim_{\Delta t \rightarrow \infty} \Big\langle W_\Sigma(r,\Delta t) \Big\rangle \ \ = \ \ A \exp\Big(-V^\Sigma(r) \Delta t\Big) \ \ = \\
\nonumber & & = \ \ A \exp\Big(-\Big(V^\textrm{$\Sigma$,(0)} + g^2 V^\textrm{$\Sigma$,(2)}(r) + 
\mathcal{O}(g^4)\Big) \Delta t\Big) \ \ = \\
\nonumber & & = \ \ A \exp\Big(-V^\textrm{$\Sigma$,(0)} \Delta t\Big) 
\exp\Big(-\Big(g^2 V^\textrm{$\Sigma$,(2)}(r) + \mathcal{O}(g^4)\Big) \Delta t\Big) \ \ = \\
\label{EQN104} & & = \ \ A \exp\Big(-V^\textrm{$\Sigma$,(0)} \Delta t\Big) 
\Big(1 - g^2 V^\textrm{$\Sigma$,(2)}(r) \Delta t + \mathcal{O}(g^4)\Big) .
\end{eqnarray}
Comparing powers of $g^2$ in (\ref{EQN103}) and (\ref{EQN104}) yields the 
singlet static potential,
\begin{eqnarray}
\label{EQN108} V^1(r) \ \ = \ \ -\frac{(N^2 - 1) g^2}{8 N \pi r} + \textrm{const} + \mathcal{O}(g^4) .
\end{eqnarray}

For the adjoint case $\Sigma = T^a$ and without summing over $a$ we have 
$\textrm{Tr}(\Sigma \Sigma^\dagger) = 1/2$ and 
$\textrm{Tr}(\Sigma T^a \Sigma^\dagger T^a) = -1/4 N$. 
Assuming exponential decay as above and comparing powers of $g^2$ one finds the result known 
in the literature,
\begin{eqnarray}
\label{EQN106} V^{T^a}(r) \ \ = \ \ +\frac{g^2}{8 N \pi r} + \textrm{const} + \mathcal{O}(g^4) .
\end{eqnarray}
Note that (\ref{EQN106}) is independent of the position of $\Sigma$, i.e.\ independent of $s$. However, since in Lorenz gauge a transfer matrix or Hamiltonian does not exist, the exponential form of (\ref{EQN104}) only holds for manifestly gauge invariant observables like the ordinary Wilson loop. By contrast, in the limit of large $\Delta t$ the correlator $\langle W_{T^a}(r,\Delta t) \rangle$ in Lorenz gauge is neither positive nor exponentially decaying. In other words, the physical meaning of the result (\ref{EQN106}), which often appears in the literature, is unclear.

The importance of a gauge-covariant time evolution of colour charges by means of parallel 
transport is highlighted by the perturbative evaluation of the string-string correlator 
(\ref{stringcor}) in Lorenz gauge, 
\begin{eqnarray}
\nonumber & & \hspace{-0.7cm} \frac{1}{N} \Big\langle \textrm{Tr}\Big(U(t_1;\bfx,\bfy) U(t_2;\bfy,\bfx)\Big) \Big\rangle_\textrm{Lorenz gauge} \ \ = \\
\nonumber & & = \ \ \frac{1}{N} \bigg\langle \textrm{Tr}\bigg(P \exp\bigg(i g \int_{-r/2}^{+r/2} dz \, A_3(-\Delta t/2,0,0,z)\bigg) \\
\nonumber & & \hspace{0.675cm} P \exp\bigg(i g \int_{+r/2}^{-r/2} dz \, A_3(+\Delta t/2,0,0,z)\bigg)\bigg) \bigg\rangle_\textrm{Lorenz gauge} \ \ = \\
\nonumber & & = \ \ 1 - \frac{(N^2 - 1) g^2}{8 N \pi^2} \bigg(\lim_{\epsilon \rightarrow 0} \bigg(+\frac{\pi r}{\epsilon} - \ln\bigg(\frac{r^2}{\epsilon^2}\bigg)\bigg) - \frac{2 r}{\Delta t} \textrm{arctan}(r / \Delta t) + \ln\bigg(\frac{\Delta t^2 + r^2}{\Delta t^2}\bigg)\bigg) \\
\nonumber & & \hspace{0.675cm} + \mathcal{O}(g^4) \ \ \stackrel{\Delta t \rightarrow \infty}{\rightarrow} \\
\label{EQN109} & & \stackrel{\Delta t \rightarrow \infty}{\rightarrow} \ \ 1 - \frac{(N^2 - 1) g^2}{8 N \pi^2} \bigg(\lim_{\epsilon \rightarrow 0} \bigg(+\frac{\pi r}{\epsilon} - \ln\bigg(\frac{r^2}{\epsilon^2}\bigg)\bigg) - \frac{r^2}{\Delta t^2}\bigg) + \mathcal{O}(g^4)
\end{eqnarray}
(cf.\ Figure~\ref{FIG002}(d)). Trying to extract the singlet potential assuming exponential behaviour of this correlator and comparing powers of $g^2$ as in (\ref{EQN104}) and (\ref{EQN109}), fails. There is no linear term in $\Delta t$, i.e.\ one obtains the physically incorrect result
\begin{eqnarray}
V^1(r) \ \ = \ \ \mathcal{O}(g^4)
\end{eqnarray}
in contrast with the string-string correlator in temporal gauge, viz.~the Wilson loop, where
the parallel transport is included.


\subsection{\label{SEC120}Gauge invariant $Q \bar{Q} Q^\textrm{ad}$ correlator}

As explained in section~\ref{SEC436}, the following gauge invariant 
correlation function is equivalent to $\langle W_{T^a}(r,\Delta t) \rangle$, when evaluated in temporal gauge for $1 \leq t_1 < t_2 < T$ (case (A)),
\begin{eqnarray}
\nonumber & & \hspace{-0.7cm} \Big\langle W_{Q \bar{Q} Q^\textrm{ad}}(r,\Delta t) \Big\rangle \ \ = \ \ \bigg\langle \frac{1}{N^2 - 1} \Big(N^2 W^L(r/2+s,\Delta t) W^R(r/2-s,\Delta t) - W_1(r,\Delta t)\Big) \bigg\rangle \\
 & &
\end{eqnarray}
(cf.\ Figure~\ref{FIG002}(c)). Evaluating this correlator in Lorenz gauge
one finds
\begin{eqnarray}
\nonumber & & \hspace{-0.7cm} \lim_{\Delta t \rightarrow \infty} \Big\langle W_{Q \bar{Q} Q^\textrm{ad}}(r,\Delta t) \Big\rangle \ \ = \\
\nonumber & & = \ \ \frac{N^2}{N^2 - 1} \bigg(1 - g^2 \bigg(V_0 - \frac{N^2 - 1}{8 N \pi (r/2+s)}\bigg) \Delta t + \mathcal{O}(g^4)\bigg) \\
\nonumber & & \hspace{0.675cm} \bigg(1 - g^2 \bigg(V_0 - \frac{N^2 - 1}{8 N \pi (r/2-s)}\bigg) \Delta t + \mathcal{O}(g^4)\bigg) \\
\nonumber & & \hspace{0.675cm} - \frac{1}{N^2 - 1} \bigg(1 - g^2 \bigg(V_0 - \frac{N^2 - 1}{8 N \pi r}\bigg) \Delta t + \mathcal{O}(g^4)\bigg) \ \ = \\
\label{EQN107} & & = \ \ 1 - g^2 \bigg(\frac{(2 N^2 - 1) V_0}{N^2 - 1} - \frac{1}{8 N \pi} \bigg(\frac{N^2}{r/2 + s} + \frac{N^2}{r/2 - s} - \frac{1}{r}\bigg)\bigg) \Delta t + \mathcal{O}(g^4)
\end{eqnarray}
(in the $W^L W^R$-term the gluon must propagate within the same loop, i.e.\ either in $W^L$ or in $W^R$; otherwise the contribution vanishes, because $\textrm{Tr}(T^a) = 0$).

As a gauge invariant observable, this loop is guaranteed to decay exponentially.
Comparing powers of $g^2$ in (\ref{EQN104}) and (\ref{EQN107}) shows that the $Q \bar{Q} Q^\textrm{ad}$ static potential is
\begin{eqnarray}
V^{Q \bar{Q} Q^\textrm{ad}}(r,s) \ \ = \ \ -\frac{g^2}{8 N \pi} \bigg(\frac{N^2}{r/2 + s} + \frac{N^2}{r/2 - s} - \frac{1}{r}\bigg) + \textrm{const} + \mathcal{O}(g^4) .
\end{eqnarray}
For $s = 0$ ($Q^\textrm{ad}$ is placed symmetrically between $Q$ and $\bar{Q}$) the result simplifies to
\begin{eqnarray}
V^{Q \bar{Q} Q^\textrm{ad}}(r,s=0) \ \ = \ \ -\frac{(4 N^2 - 1) g^2}{8 N \pi r} + \textrm{const} + \mathcal{O}(g^4) ,
\end{eqnarray}
i.e.\ is in leading non-trivial order attractive and, depending on $N$, by a factor $4 \ldots 5$ stronger than the singlet static potential (\ref{EQN108}).


\subsection{Coulomb gauge}

The gluon propagator in Coulomb gauge $\partial_j A_j = 0$ is
\begin{eqnarray}
 & & \hspace{-0.7cm} \tilde{D}_{F,0 0}^{a b}(p) \ \ = \ \ \delta^{a b} \frac{1}{|\mathbf{p}|^2} \quad , \quad \tilde{D}_{F,j k}(p) \ \ = \ \ \delta^{a b} \frac{1}{p^2} \bigg(\delta_{j k} - \frac{p_j p_k}{|\mathbf{p}|^2}\bigg) \\
 & & \hspace{-0.7cm} D_{F,0 0}^{a b}(x-y) \ \ = \ \ \frac{1}{(2 \pi)^4} \int d^4p \, e^{-i p (x-y)} \tilde{D}_{F,0 0}^{a b}(p) \ \ = \ \ \delta^{a b} \frac{\delta(x_0 - y_0)}{4 \pi |\mathbf{x} - \mathbf{y}|} .
\end{eqnarray}


Starting from (\ref{pert}) the spatial parallel transporters can again be neglected for $\Delta t \rightarrow \infty$, while the integrals along the temporal lines give
\begin{eqnarray}
 & & \hspace{-0.7cm} \lim_{\epsilon \rightarrow 0} \int_{-\Delta t/2}^{+\Delta t/2} dt_1 \, \int_{-\Delta t/2}^{+\Delta t/2} dt_2 \, \frac{\delta(t_1 - t_2)}{4 \pi \epsilon} \ \ = \ \ \frac{\Delta t}{4 \pi \epsilon} \\
 & & \hspace{-0.7cm} \int_{-\Delta t/2}^{+\Delta t/2} dt_1 \, \int_{+\Delta t/2}^{-\Delta t/2} dt_2 \, \frac{\delta(t_1 - t_2)}{4 \pi r} \ \ = \ \ -\frac{\Delta t}{4 \pi r} .
\end{eqnarray}
The result for $\langle W_\Sigma(r,\Delta t) \rangle$ is is the same as in Lorenz gauge. Consequently,
\begin{eqnarray}
V^1(r) \ \ = \ \ -\frac{(N^2 - 1) g^2}{8 N \pi r} + \textrm{const} + \mathcal{O}(g^4)
\end{eqnarray}
is the same in both gauges, which is expected, since the Wilson loop $W_1(r,\Delta t)$ is gauge invariant. Similarly one finds the same result for the colour-adjoint static potential obtained from $W_{T^a}(r,\Delta t)$ in Coulomb gauge as well as in Lorenz gauge,
\begin{eqnarray}
\label{EQN110} V^{T^a}(r) \ \ = \ \ +\frac{g^2}{8 N \pi r} + \textrm{const} + \mathcal{O}(g^4) .
\end{eqnarray}
In Coulomb gauge a transfer matrix/Hamiltonian exists \cite{Zwanziger:1995cv}, the asymptotic decay in $t$ is exponential and a comparison of 
powers of $g^2$ in (\ref{EQN104}) and $W^{T^a}$ (which yields (\ref{EQN110})) 
appears physically meaningful. However, a non-zero result without completing the
gauge is in contrast to the non-perturbative evaluation, section~\ref{sec:coul}.
This indicates that gauge fixing in perturbation theory is not the same 
as in a corresponding non-perturbative formulation. 
The reason is that perturbation theory is an expansion around $A_\mu = 0$, 
which does not include averaging over all gauge equivalent 
gauge field configurations fulfilling the gauge condition.


\section{\label{SEC500}Numerical results}

In this section we compute the singlet static potential $V^1(r)$ and the colour-adjoint static potential in temporal gauge $V^{T^a}(r)$, which is identical to the gauge invariant static potential $V^{Q \bar{Q} Q^\textrm{ad}}$, using $SU(2)$ lattice gauge theory, where $T^a = \sigma^a/2$.


\subsection{Lattice setup}

The lattice action is the standard Wilson plaquette gauge action (eqs.\ (\ref{EQN468}) to (\ref{EQN469})). The lattice extension is $(L/a)^3 \times T/a = 24^3 \times 48$. We have performed simulations at four different values of the coupling constant $\beta \in \{ 2.40 , 2.50 , 2.60 , 2.70 \}$. Observables are computed as averages over 200 essentially independent gauge link configurations.

 To introduce a physical scale we have identified the Sommer parameter $r_0$ with $0.46 \, \textrm{fm}$, which is roughly the QCD value (cf.\ e.g.\ \cite{Baron:2010bv,Donnellan:2010mx}; the Sommer parameter is defined via the static force, $|F(r_0) r_0^2| \equiv 1.65$, $F(r) = d V^1(r) / dr$). The corresponding lattice spacings are listed in Table~\ref{TAB001}.

\begin{table}[htb]
\begin{center}
\begin{tabular}{|c|c|c|c|c|}
\hline
 & & & & \vspace{-0.40cm} \\
$\beta$ & $a$ in $\textrm{fm}$ & $\alpha_s^1$ & $\alpha_s^{T^a}$ & $\Delta \alpha_s^\textrm{rel}$ \\
 & & & & \vspace{-0.40cm} \\
\hline
 & & & & \vspace{-0.40cm} \\
$2.40$ & $0.102$ & $0.89$ & $0.75$ & $17 \%$ \\
$2.50$ & $0.073$ & $0.59$ & $0.52$ & $13 \%$ \\
$2.60$ & $0.050$ & $0.43$ & $0.40$ & $\phantom{0}9 \%$ \\
$2.70$ & $0.038$ & $0.36$ & $0.33$ & $\phantom{0}6 \%$\vspace{-0.40cm} \\
 & & & & \\
\hline
\end{tabular}
\caption{\label{TAB001}$\alpha_s$ extracted from $V^1(r)$ and from $V^{T^a}(r)$.}
\end{center}
\end{table}






\subsection{The singlet and the colour-adjoint/$Q \bar{Q} Q^\textrm{ad}$ static potential}

We have used standard techniques, to compute the singlet and the colour-adjoint/$Q \bar{Q} Q^\textrm{ad}$ static potential. We have evaluated the gauge invariant correlator diagrams shown in Figure~\ref{FIG003} (``singlet, case (A)'' and ``adjoint, case (A)'') on the 200 available gauge link configurations using translational and rotational invariance to increase statistical precision. Moreover, we have resorted to common smearing techniques:
\begin{itemize}
\item[(1)] \textbf{APE smearing for spatial links} \cite{Albanese:1987ds}: \\ This improves the ground state overlap, i.e.\ increases e.g.\ for the matrix elements in (\ref{EQN210}) the ratio
\begin{eqnarray}
\left|\frac{\langle 0_{\alpha \beta} | \hat{U}_{\alpha \beta}(\bfx,\bfy) | 0 \rangle}{\langle k'_{\alpha \beta} | \hat{U}_{\alpha \beta}(\bfx,\bfy) | n' \rangle}\right| \quad , \quad (k',n') \ \ \neq \ \ (0,0) .
\end{eqnarray}
This in turn allows, to extract the value of $V^1(r)$ or $V^{T^a}(r)$ from the exponential behavior of the corresponding correlator diagrams at smaller temporal separations, where also the statistical errors are smaller. We used $N_\textrm{APE} = 15$ and $\alpha_\textrm{APE} = 0.5$, where detailed equations can e.g.\ be found in \cite{Jansen:2008si}, section~3.1.

\item[(2)] \textbf{HYP2 smearing for temporal links} \cite{Hasenfratz:2001hp,DellaMorte:2003mn,Della Morte:2005yc}: \\ This amounts to using a different discretisation for the action of the static colour charges. It reduces the self energy of these charges and, therefore, the extracted static potential $V^1(r)$ or $V^{T^a}(r)$ by an $r$-independent constant. Consequently, the exponential decay of the correlator diagram is weaker, which again results in a smaller statistical error.
\end{itemize}

\begin{figure}[htb]
\begin{center}
\includegraphics[angle=-90,width=0.80\textwidth]{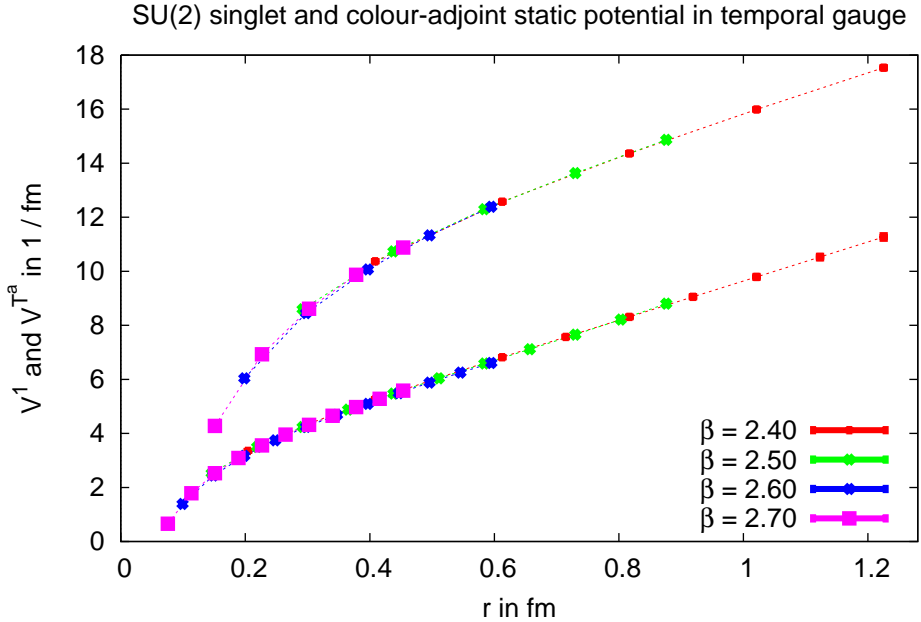}
\caption{\label{FIG004}$V^1$ (lower curve) and $V^{T^a}$ (upper curve) as a functions of the $Q \bar{Q}$ separation $r$ in physical units (lattice results obtained at different values of $\beta$ have been shifted vertically, to compensate for the $a$-dependent self energy of the static charges). For $V^{T^a}$ the adjoint static colour charge is located halfway between $Q$ and $\bar{Q}$. Lattice results, where any of the static charges are closer than $2 a$, have been omitted, because of rapidly increasing discretisation errors.}
\end{center}
\end{figure}

Figure~\ref{FIG004} shows $V^1$ and $V^{T^a}$ in physical units. Note that at large static colour charge separations both $V^1$ and $V^{T^a}$ exhibit essentially the same slope. This is expected and indicates flux tube formation not only between $Q$ and $\bar{Q}$ in the singlet case, but also between $Q$ and $Q^\textrm{ad}$ and also between $Q^\textrm{ad}$ and $\bar{Q}$ in the colour-adjoint/$Q \bar{Q} Q^\textrm{ad}$ case.

\subsection{Comparing lattice and perturbative results}

Lattice results for the static potential are known to exhibit large discretisation errors, as soon as static colour charges are closer than $2 a$ (for our ensembles $2 a \approx 0.08 \, \textrm{fm} \ldots 0.20 \, \textrm{fm}$). On the other hand perturbative results for the static potential are only trustworthy for separations $\ltapprox 0.2 \, \textrm{fm}$. Therefore, the region of overlap between lattice and perturbative results is quite small. Moreover, the leading order of perturbation theory
\begin{eqnarray}
V^\textrm{1,\textrm{LO}}(r) \ \ = \ \ -\frac{3 g^2}{16 \pi r} + \textrm{const} \quad , \quad V^{T^a,\textrm{LO}}(r) \ \ = \ \ -\frac{15 g^2}{16 \pi r} + \textrm{const}
\end{eqnarray}
(here specialized to gauge group $SU(2)$) we have calculated in sections~\ref{SEC119} and \ref{SEC120} is known to be a rather poor approximation (cf.\ e.g.\ \cite{Brambilla:2010pp,Jansen:2011vv,Bazavov:2012ka}). Consequently, one can only expect qualitative agreement, when comparing the here presented lattice and perturbative results.

We perform such a comparison by determining $\alpha_s \equiv g^2 / 4 \pi$ from the corresponding static forces $F^X(r) = dV^X(r) / dr$, $X \in \{ 1 , T^a \}$, which we define on the lattice by finite differences. For the singlet case we use
\begin{eqnarray}
\frac{V^{1,\textrm{lattice}}(3a) - V^{1,\textrm{lattice}}(2a)}{a} \ \ = \ \ \frac{3 \alpha_s^1}{4 (2.5 \times a)^2} .
\end{eqnarray}
This ensures that the static colour charges are separated by at least $2 a$, while at the same time their separation is small enough, to expect that the leading order perturbative result is a reasonable approximation. Similarly we use
\begin{eqnarray}
\frac{V^{T^a,\textrm{lattice}}(6a) - V^{T^a,\textrm{lattice}}(4a)}{2 a} \ \ = \ \ \frac{15 \alpha_s^{T^a}}{4 (5 \times a)^2}
\end{eqnarray}
for the colour-adjoint case.

The values we for $\alpha_s$ we obtain are collected in Table~\ref{TAB001}. The relative difference between $\alpha_s$ extracted from $V^1(r)$ and from $V^{T^a}(r)$, defined as
\begin{eqnarray}
\Delta \alpha_s^\textrm{rel} \ \ \equiv \ \ 2 \left|\frac{\alpha_s^1 - \alpha_s^{T^a}}{\alpha_s^1 + \alpha_s^{T^a}}\right| ,
\end{eqnarray}
is quite small, less than $10 \%$ for our two smallest lattice spacings, which is a clear sign of agreement between the lattice and the perturbative results. $\Delta \alpha_s^\textrm{rel}$ is getting smaller, when the lattice spacing is decreased, which is expected, since the quality of the leading order perturbative approximation is improving at smaller static colour charge separations.

It is also interesting to note that the values for $\alpha_s$ increase for larger lattice spacings, i.e.\ for larger static colour charge separations. $\alpha_s \gtapprox 1$ signals complete breakdown of perturbation theory.


\section{Conclusions}

We have discussed the non-perturbative definition of a static
potential for a quark anti-quark pair in a colour-adjoint configuration, based on Wilson loops
with generator insertions in the spatial string. Leading order perturbation theory
in Lorenz gauges has long predicted the corresponding potential to be repulsive. 
Saturating the open adjoint indices with
colour-magnetic fields, as suggested in the literature \cite{pnrqcd}, produces a well defined
gauge invariant observable but spectral analysis shows it to project on the colour-singlet
channel only. If the adjoint indices are left open, gauge fixing is required in order to obtain
a non-zero result for the correlator. 

Lorenz gauges violate positivity and the gauge fixed correlator no longer
has purely exponential decay, thus precluding a non-perturbative definition of the potential.
In temporal gauge a positive transfer matrix with 
well defined charge sectors exists and the Wilson loop with generator insertions can be 
shown to be equivalent
to a gauge invariant object, where the open charges are saturated with an adjoint 
Schwinger line. This correlator projects on states with the desired
transformation behaviour, however the resulting potential is attractive and should be  
interpreted as a three quark potential (fundamental, anti-fundamental and adjoint).
The same qualitative behaviour is found in leading order perturbation theory once the adjoint line
is included.
In Coulomb gauge a positive transfer matrix exists, but the gauge is incomplete and the 
observable averages to zero. Imposing an additional gauge condition to render Coulomb gauge complete again introduces an adjoint static quark. 
The interpretation of the resulting static potential is then identical to that in temporal gauge.
It thus appears impossible to reproduce the perturbatively repulsive colour-adjoint potential by a non-perturbative 
computation based on Wilson loops, even at short distance. 

In a recent similar work a non-perturbative extraction of the colour-adjoint potential 
from Polyakov loop correlators was suggested \cite{rossi}. Similar to our treatment here and in earlier work \cite{Wagner:2012pk}, an adjoint
Schwinger line appears, which however is placed at spatial infinity, or far away from 
the fundamental quarks. While no simulations of this observable are available yet, since 
adjoint charges are screened we expect this correlator
to decay as the ordinary Polyakov loop correlator with the singlet potential shifted by a gluelump mass.

\section*{Acknowledgments}

We thank Felix Karbstein for discussions.

M.W.\ acknowledges support by the Emmy Noether Programme of the DFG (German Research Foundation), grant WA 3000/1-1.

This work was supported in part by the Helmholtz International Center for FAIR within the framework of the LOEWE program launched by the State of Hesse.




\begin{thebibliography}{99}

\bibitem{nrqcd}
G.~T.~Bodwin, E.~Braaten and G.~P.~Lepage,
  Phys.\ Rev.\ D {\bf 51} (1995) 1125
   [Erratum-ibid.\ D {\bf 55} (1997) 5853]
  [hep-ph/9407339].

\bibitem{pnrqcd}
  N.~Brambilla, A.~Pineda, J.~Soto and A.~Vairo,
  ``Potential NRQCD: an effective theory for heavy quarkonium,''
  Nucl.\ Phys.\ B {\bf 566}, 275 (2000)
  [hep-ph/9907240].

\bibitem{brown}
L.~S.~Brown and W.~I.~Weisberger,
  Phys.\ Rev.\ D {\bf 20} (1979) 3239.

\bibitem{svet}
L.~D.~McLerran and B.~Svetitsky,
  Phys.\ Rev.\ D {\bf 24} (1981) 450.

\bibitem{nad}
S.~Nadkarni,
  Phys.\ Rev.\ D {\bf 34} (1986) 3904.

\bibitem{shur}
E.~V.~Shuryak and I.~Zahed,
  Phys.\ Rev.\ D {\bf 70} (2004) 054507
  [hep-ph/0403127].

\bibitem{pet}
N.~Brambilla, J.~Ghiglieri, A.~Vairo and P.~Petreczky,
  Phys.\ Rev.\ D {\bf 78} (2008) 014017
  [arXiv:0804.0993 [hep-ph]].

\bibitem{kniehl}
B.~A.~Kniehl, A.~A.~Penin, Y.~Schroder, V.~A.~Smirnov and M.~Steinhauser,
  Phys.\ Lett.\ B {\bf 607} (2005) 96
  [hep-ph/0412083].

\bibitem{jp}
  O.~Jahn and O.~Philipsen,
  ``The Polyakov loop and its relation to static quark potentials and free energies,''
  Phys.\ Rev.\ D {\bf 70}, 074504 (2004)
  [hep-lat/0407042].

\bibitem{seiler}
C.~Borgs and E.~Seiler,
  Commun.\ Math.\ Phys.\  {\bf 91} (1983) 329.

\bibitem{Zwanziger:1995cv} 
  D.~Zwanziger,
  Nucl.\ Phys.\ B {\bf 485}, 185 (1997)
  [hep-th/9603203].

\bibitem{Wagner:2012pk} 
  M.~Wagner and O.~Philipsen,
  arXiv:1211.2165 [hep-lat].

\bibitem{Philipsen:2001ip}
  O.~Philipsen,
  ``On the nonperturbative gluon mass and heavy quark physics,''
  Nucl.\ Phys.\ B {\bf 628}, 167 (2002)
  [hep-lat/0112047].

\bibitem{Montvay:1994cy} 
  I.~Montvay and G.~M\"unster,
  ``Quantum fields on a lattice,''
  Cambridge University Press (1994).

\bibitem{Creutz:1978ub} 
  M.~Creutz,
  ``On invariant integration over SU(N),''
  J.\ Math.\ Phys.\ {\bf 19}, 2043 (1978).

\bibitem{Bali:2005fu} 
  G.~S.~Bali {\it et al.} [SESAM Collaboration],
  ``Observation of string breaking in QCD,''
  Phys.\ Rev.\ D {\bf 71}, 114513 (2005)
  [hep-lat/0505012].

\bibitem{Wagner:2010ad} 
  M.~Wagner [ETM Collaboration],
  ``Forces between static-light mesons,''
  PoS LATTICE {\bf 2010}, 162 (2010)
  [arXiv:1008.1538 [hep-lat]].

\bibitem{Baron:2010bv} 
  R.~Baron {\it et al.} [ETM Collaboration],
  ``Light hadrons from lattice QCD with light $(u,d)$, strange and charm dynamical quarks,''
  JHEP {\bf 1006}, 111 (2010)
  [arXiv:1004.5284 [hep-lat]].

\bibitem{Donnellan:2010mx} 
  M.~Donnellan, F.~Knechtli, B.~Leder and R.~Sommer,
  ``Determination of the static potential with dynamical fermions,''
  Nucl.\ Phys.\ B {\bf 849}, 45 (2011)
  [arXiv:1012.3037 [hep-lat]].

\bibitem{Albanese:1987ds}
  M.~Albanese {\it et al.} [APE Collaboration],
  ``Glueball masses and string tension in lattice QCD,''
  Phys.\ Lett.\ B {\bf 192}, 163 (1987).

\bibitem{Jansen:2008si} 
  K.~Jansen {\it et al.} [ETM Collaboration],
  ``The Static-light meson spectrum from twisted mass lattice QCD,''
  JHEP {\bf 0812}, 058 (2008)
  [arXiv:0810.1843 [hep-lat]].

\bibitem{Hasenfratz:2001hp}
  A.~Hasenfratz and F.~Knechtli,
  ``flavour symmetry and the static potential with hypercubic blocking,''
  Phys.\ Rev.\ D {\bf 64}, 034504 (2001)
  [arXiv:hep-lat/0103029].

\bibitem{DellaMorte:2003mn}
  M.~Della Morte {\it et al.},
  ``Lattice HQET with exponentially improved statistical precision,''
  Phys. Lett. {\bf B581}, 93, (2004)
  [arXiv:hep-lat/0307021].

\bibitem{Della Morte:2005yc}
  M.~Della Morte, A.~Shindler and R.~Sommer,
  ``On lattice actions for static quarks,''
  JHEP {\bf 0508}, 051 (2005)
  [arXiv:hep-lat/0506008].

\bibitem{Brambilla:2010pp} 
  N.~Brambilla, X.~Garcia i Tormo, J.~Soto and A.~Vairo,
  ``Precision determination of $r_0 \Lambda_{\overline{\rm MS}}$ from the QCD static energy,''
  Phys.\ Rev.\ Lett.\ {\bf 105}, 212001 (2010)
  [Erratum-ibid.\ {\bf 108}, 269903 (2012)]
  [arXiv:1006.2066 [hep-ph]].

\bibitem{Jansen:2011vv}
  K.~Jansen {\it et al.} [ETM Collaboration],
  ``$\Lambda_{\bar{MS}}$ from the static potential for QCD with $n_f=2$ dynamical quark flavors,''
  JHEP {\bf 1201}, 025 (2012)
  [arXiv:1110.6859 [hep-ph]].

\bibitem{Bazavov:2012ka}
  A.~Bazavov, N.~Brambilla, X.~Garcia i Tormo, P.~Petreczky, J.~Soto and A.~Vairo,
  ``Determination of $\alpha_s$ from the QCD static energy,''
  arXiv:1205.6155 [hep-ph].

\bibitem{rossi}
G.~Rossi and M.~Testa,
  Phys.\  Rev.\  D 87, {\bf 085014} (2013)
  [arXiv:1304.2542 [hep-lat]].

\end{thebibliography}
\end{document}